\documentclass[preprint]{aastex}
\slugcomment{To appear in PASP (2005, v. 117, p. 445)}
\shorttitle{Theoretical Isochrones near the K Band}
\shortauthors{KIM, FIGER, LEE, \& OH}
\def\spose#1{\hbox to 0pt{#1\hss}}
\newcommand\lsim{\mathrel{\spose{\lower 3.0pt\hbox{$\mathchar"218$}}
     \raise 2.0pt\hbox{$\mathchar"13C$}}}
\newcommand\gsim{\mathrel{\spose{\lower 3.0pt\hbox{$\mathchar"218$}}
     \raise 2.0pt\hbox{$\mathchar"13E$}}}

\begin{document}
\title{Theoretical Isochrones with Extinction in the K-Band}
\author{Sungsoo S. Kim\altaffilmark{1}, Donald F. Figer\altaffilmark{2},
Myung Gyoon Lee\altaffilmark{3}, and Seungkyung Oh\altaffilmark{1}}
\altaffiltext{1}{Dept. of Astronomy \& Space Science, Kyung Hee University,
Yongin-shi, Kyungki-do 449-701, Korea; sskim@ap.khu.ac.kr}
\altaffiltext{2}{Space Telescope Science Institute, 3700 San Martin Drive,
Baltimore, MD 21218; figer@stsci.edu}
\altaffiltext{3}{Astronomy Program, SEES, Seoul National University,
Seoul 151-742, Korea; mglee@astrog.snu.ac.kr}

%%%%%%%%%%%%%%%%%%%%%%%%%%%%%%%%%%%%%%%%%%%%%%%%%%%%%%%%%%%%%%%%%%%%%%%%%%%%%%%%
\begin{abstract}
We calculate theoretical isochrones, in a consistent way, for five
filters in the atmospheric window between $1.9 \, {\rm \mu m}$ and
$2.5 \, {\rm \mu m}$, $K$, $K'$, $K_s$, F205W, and F222M, using
the Padova stellar evolutionary models by Girardi et al.
Even when displayed in the same Vega magnitude system, the near-infrared
colors of the same isochrone can differ by up to 0.18 mag at its
bright end, depending on the filter.  We present magnitude transformations
between $K$-band filters as a function of color from $H$ \& $K$
band filters. Isochrones with extinction at $K$
of up to 6 mag are also presented.  We find that care is needed
when comparing extinction values that are estimated using different
filter sets in the $K$-band, in particular when comparing those
between atmospheric and space filter sets: extinction values for
space filters can be in error by up to 0.3 mag.
To reduce this error, we introduce an ``effective extinction slope''
for each filter set and isochrone model, which describes the extinction
behaviour of isochrones in the color-magnitude diagram more correctly
than the actual extinction law.  Our calculation also suggests that
the extinction law implied by the observations of Rieke, Rieke, \& Paul
for wavelengths between $H$ and $K$ bands is better described by a
power-law function with an exponent of 1.61, instead of 1.55, which is
commonly used with an assumption that the transmission functions of $H$
and $K$ filters are Dirac delta functions.
\end{abstract}

%%%%%%%%%%%%%%%%%%%%%%%%%%%%%%%%%%%%%%%%%%%%%%%%%%%%%%%%%%%%%%%%%%%%%%%%%%%%%%%%
\section{INTRODUCTION}
\label{sec:introduction}

Deep near-infrared stellar photometry is an important tool for studies
of the structure and stellar population in the Galactic bulge and the disk,
where the extinction by dust is quite significant.
In the near-infrared ($1-5 \, \mu$m), deep ground-based 
imaging is generally limited to the three
shorter atmospheric windows, $J$ ($1.2 \, \mu$m), $H$ ($1.65 \, \mu$m),
and $K$ ($2.2 \, \mu$m). Longward of $2.5 \, \mu$m,
thermal emission from a warm telescope, instrument, and sky (for ground-based
observations) makes photometry highly inefficient.

The $K$-band is the most popular among these three bands, because
it suffers from the least extinction and provides the widest wavelength
baseline when combined with visible bands.  Contamination
by the thermal emission in the $K$ filter (Johnson et al. 1966) is not
negligible, so astronomers have introduced several filters with slightly
shorter long wavelength cutoffs,
i.e.\ $K'$ (Wainscoat \& Cowie 1992) and $K_s$ ($K$-short; developed by
M. Skrutskie---see the appendix of Persson et al. 1998), that effectively
reduce the total background from a warm sky and telescope by a factor of two.

On the other hand, the near-infrared camera on board {\it Hubble Space
Telescope} ({\it HST}), NICMOS, has a unique filter system, as it is
not subject to the effects of atmospheric absorption.  The NICMOS F160W
filter and the $H$ filter (Johnson et al. 1966) have similar central
wavelengths, and the NICMOS F205W and F222M filters transmit wavelengths that
fall within the passband of the Johnson $K$ filter.

\begin{figure}
%Fig 1
\epsscale{0.9}
\plotone{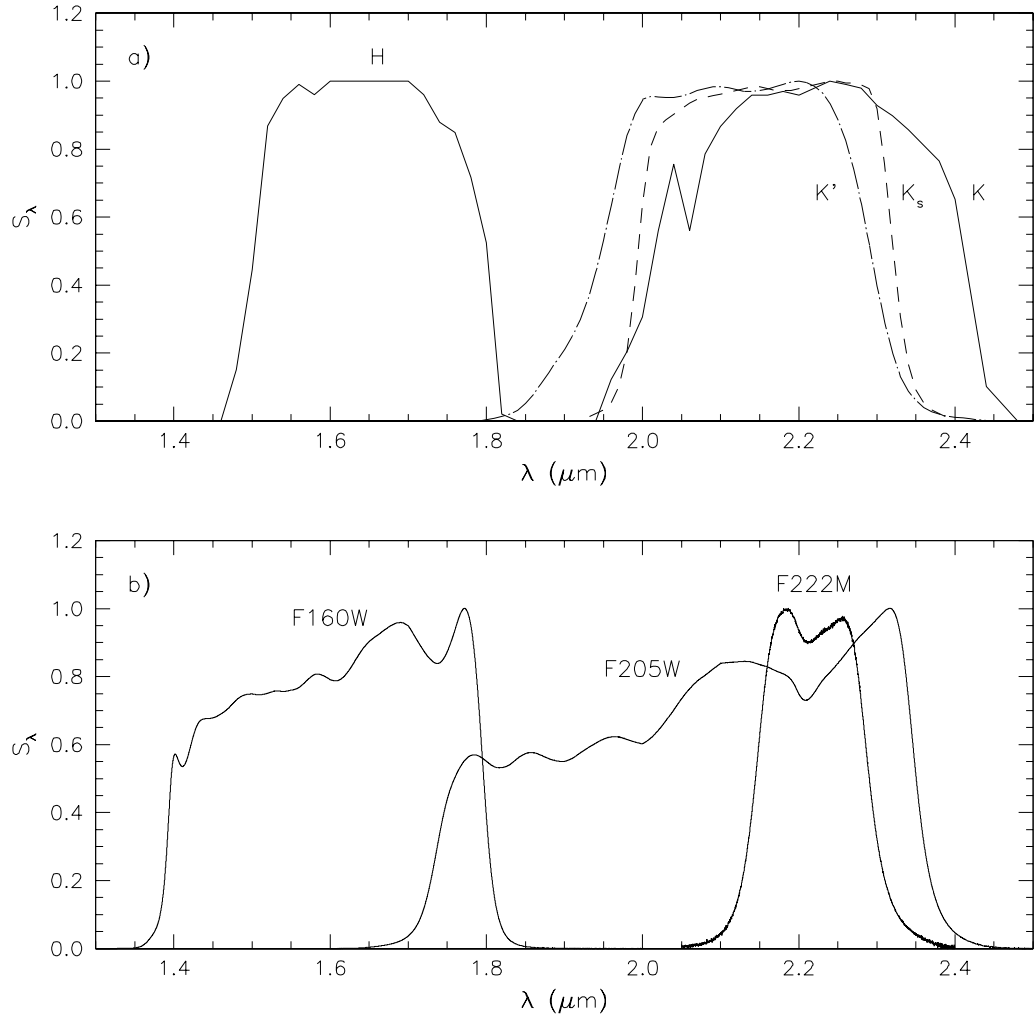}
\caption
{\label{fig:filters}Transmission functions ($S_\lambda$) of the filters
considered in the present work.  $S_\lambda$'s are scaled such that their
maximum values become 1.}
\end{figure}

As enumerated above, there are at least five different broad- and medium-band
filters in common usage that cover wavelengths near $2.2 \, {\rm \mu m}$: 
three atmospheric filters $K$, $K'$, \& $K_s$, and
two space filters F205W \& F222M.  Transmission functions for 
these five filters are quite different from each other (see
Figure~\ref{fig:filters}), but there have been no studies that consistently
compare the magnitudes and colors of stars as observed with these five filters
as a function of interstellar extinction.

There have been a few studies that compare theoretical isochrones or colors for
{\it some} of the five filters.  Girardi et al.\ (2002) presented theoretical
isochrones for $K$, $K_s$, and F205W filters among others, but not for $K'$
and F222M.  Origlia \& Leitherer (2000) provided color transformations
between $J-H$ \& F110W$-$F160W, and between $J-K$ \& F110W$-$F222M, and
Stephens et al.\ (2000) presented transformation between $K$ magnitude and
F110W$-$F205W color, but none for $K'$, $K_s$, or F205W.
Wainscoat \& Cowie (1992) investigated the transformation between the
$K'$ and $K$ filters.

In the present paper, we present and analyze theoretical isochrones 
that are calculated in a consistent way for the five filters
near the $2 \, {\rm \mu m}$ atmospheric window over a range of extinctions.

The magnitude system we adopt here is explained in \S~\ref{sec:magsys},
and the stellar spectral library and evolutionary tracks we use are
discussed in \S~\ref{sec:spec_track}.  Theoretical isochrones are
presented in \S~\ref{sec:isochrones}, and our findings are then summarized
in \S~\ref{sec:summary}.

%%%%%%%%%%%%%%%%%%%%%%%%%%%%%%%%%%%%%%%%%%%%%%%%%%%%%%%%%%%%%%%%%%%%%%%%%%%%%%%%
\section{Magnitude System}
\label{sec:magsys}

Here, we briefly discuss our adopted magnitude system, 
following the discussion by Girardi et al. (2002).

The flux of a star measured at the top of the Earth's atmosphere, $f_\lambda$,
is related to the flux at the stellar photosphere, $F_\lambda$, by
\begin{equation}
	f_\lambda = 10^{-0.4 A_\lambda} (R/d)^2 F_\lambda,
\end{equation}
where $A_\lambda$ is the extinction in magnitudes at the wavelength
$\lambda$, $R$ is the stellar radius, and $d$ is the distance to the star.
The apparent magnitude, $m_{S_\lambda}$, as measured using a filter with
transmission function $S_\lambda$ is given by
\begin{equation}
\label{mag}
	m_{S_\lambda} = -2.5 \log \left (
                        \frac{\int_{\lambda 1}^{\lambda 2} f_\lambda
                              S_\lambda d \lambda}
                             {\int_{\lambda 1}^{\lambda 2} f_\lambda^0
                              S_\lambda d \lambda}
			\right ) + m_{S_\lambda}^0,
\end{equation}
where $f_\lambda^0$ is the reference spectrum corresponding to a known apparent
magnitude $m_{S_\lambda}^0$. The integrands of equation~(\ref{mag})
correspond to the total photon energy, which suitably represents
data obtained with traditional energy-collecting devices, such as bolometers.
However, data from modern photon-collecting
devices, such as CCDs and infrared arrays, would be better represented
by an integration over the collected photon flux,
\begin{equation}
\label{mag_ph}
	m_{S_\lambda} = -2.5 \log \left (
                        \frac{\int_{\lambda 1}^{\lambda 2} \lambda f_\lambda
                              S_\lambda d \lambda}
                             {\int_{\lambda 1}^{\lambda 2} \lambda f_\lambda^0
                              S_\lambda d \lambda}
			\right ) + m_{S_\lambda}^0.
\end{equation}
The difference between energy integration and photon integration is
usually very small when extinction is negligible.  In the present paper,
we adopt the photon integration (eq. \ref{mag_ph}) for defining the magnitudes.

By combining the equations above, we derive the absolute magnitude,
for which $d$ is defined to be 10~pc:
\begin{equation}
\label{Mag}
	M_{S_\lambda} = -2.5 \log \left [
			\left ( \frac{R}{10{\rm pc}}  \right )^2
                        \frac{\int_{\lambda 1}^{\lambda 2} 
			      10^{-0.4 A_\lambda} \lambda F_\lambda S_\lambda
                              d \lambda}
                             {\int_{\lambda 1}^{\lambda 2} \lambda f_\lambda^0
                              S_\lambda d \lambda}
			\right ] + m_{S_\lambda}^0.
\end{equation}
Stellar radius $R$ can be obtained if the luminosity $L$ and
the effective surface temperature $T_{eff}$ of the star are given.

We adopt a Vega-based photometric system (VEGAmag system), which uses Vega
($\alpha$ Lyr) as the calibrating star.  Following Girardi et al. (2002),
we adopt the synthetic ATLAS9 model (Kurucz 1993) for Vega with $T_{eff} =
9550$~K, $\log g = 3.95$, ${\rm [M/H]} = -0.5$, microturbulent velocity
$\xi = 2 \, {\rm km/s}$, $(R/d)^2 = 6.247 \times 10^{-17}$, and photometric
zeropoint of 0.03.  The dilution factor $(R/d)^2$ is needed to calculate
$f_\lambda^{Vega}$ from the model spectrum $F_\lambda^{Vega}$.  Now,
equation~(\ref{Mag}) can be written as
\begin{equation}
\label{Mag2}
	M_{S_\lambda} = -2.5 \log \left [
                        \frac{(R/10{\rm pc})^2 \int_{\lambda 1}^{\lambda 2} 
			      10^{-0.4 A_\lambda} \lambda F_\lambda S_\lambda
                              d \lambda}
                             {(R_{\rm Vega}/d_{\rm Vega})^2
			      \int_{\lambda 1}^{\lambda 2} \lambda
			      F_\lambda^{\rm Vega} S_\lambda d \lambda}
			\right ] + m_{S_\lambda}^{\rm Vega},
\end{equation}
where $(R_{\rm Vega}/d_{\rm Vega})^2 = 6.247 \times 10^{-17}$ and
$m_{S_\lambda}^{\rm Vega} = 0.03$.

The NICMOS Data Handbook (Dickinson 2002)
defines the Vega magnitude system for NICMOS filters
by using a filter-averaged flux density of a synthetic model for Vega,
$\langle f_\nu^{\rm Vega} \rangle$, as a photometric zeropoint.  Therefore,
for NICMOS filters, the magnitude is defined to be
\begin{equation}
\label{Mag2NIC}
	M_{S_\lambda} = -2.5 \log \left [
                        \frac{(R/10{\rm pc})^2 \int_{\lambda 1}^{\lambda 2} 
			      10^{-0.4 A_\lambda} \lambda F_\lambda S_\lambda
                              d \lambda}
			     {\langle f_\nu^{\rm Vega} \rangle
                              c \int_{\nu 1}^{\nu 2} \nu^{-1} \lambda S_\nu
                              d \nu}
			\right ] + m_{S_\lambda}^{\rm Vega}.
\end{equation}
We adopt $\langle f_\nu^{\rm Vega} \rangle$ values from the NICMOS Data Handbook
Ver. 5.0: 1040.7~Jy for F160W, 703.6~Jy for F205W, and 610.4~Jy for F222M.

We adopt filter transmission functions of Bessell \& Brett (1988) for $H$ and
$K$, and those of the Quirk camera of the University of Hawaii for $K'$ and
$K_s$.\footnote{http://www.ifa.hawaii.edu/instrumentaion/quirc/quirc.html}
In the case of NICMOS filters, transmission functions include both throughputs
and detector quantum efficiencies, which come from SYNPHOT, a synthetic
photometry package distributed as a part of STSDAS (Space Telescope
Science Data Analysis System) software.  When calculating the magnitudes
for ground-based filters, we do not consider the atmospheric transmission,
telescope throughputs, and detector quantum efficiencies with an assumption
that the wavelegnth dependence of these effects is negligible for the
wavelength ranges covered by the filters considered here.

%%%%%%%%%%%%%%%%%%%%%%%%%%%%%%%%%%%%%%%%%%%%%%%%%%%%%%%%%%%%%%%%%%%%%%%%%%%%%%%%
\section{Stellar Spectral Library \& Evolutionary Tracks}
\label{sec:spec_track}

For the spectra of synthetic stellar atmospheres, we adopt Kurucz
ATLAS9 no-overshoot models\footnote{``NOVER" files at
http://kurucz.harvard.edu/grids.html} calculated by Castelli et al. (1997).
The metallicities of these models cover the values of [M/H] = $-2.5$ to $+0.5$.
A microturbulent velocity $\xi=2\,{\rm km/s}$, and a mixing length
parameter $\alpha =1.25$ are adopted in the present study.

For the temporal evolution of $T_{eff}$ and $L$ as a function
of stellar mass, i.e.\ stellar evolutionary tracks, we adopt the ``basic set" of
the Padova models\footnote{http://pleiadi.pd.astro.it} (Girardi et al. 2002).
We consider isochrones with a metallicity $Z$ = 0.0001, 0.001, 0.019, \& 0.03.
The stellar spectral library and the evolutionary tracks we adopted assume
a solar chemical mixture.

\begin{figure}
%Fig 2
\epsscale{0.6}
\plotone{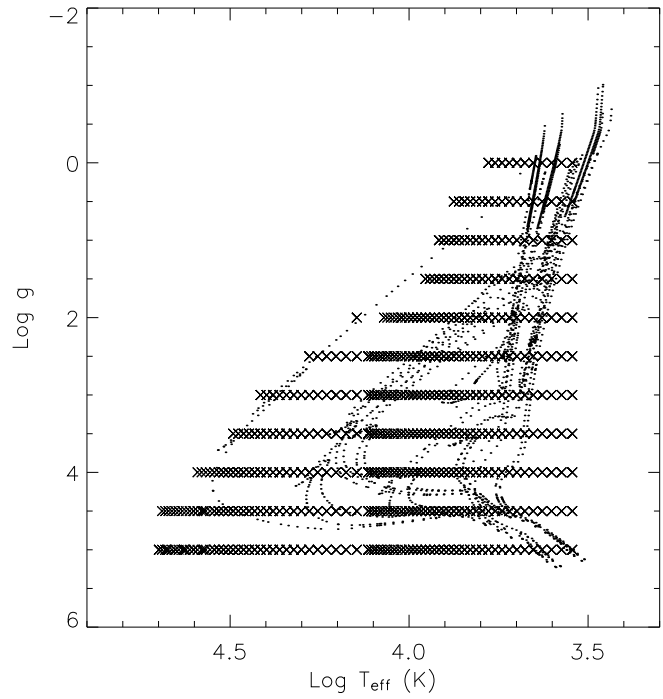}
\caption
{\label{fig:tg}[$\log T_{eff}$, $\log g$] distribution of the spectra
that we use in the present work among the stellar spectral library 
by Castelli et al. (1997; crosses), and the same distribution of the
stellar models that we use among the stellar evolutionary tracks by
Girardi et al. (2002; dots).}
\end{figure}

ATLAS9 models cover a wide range of $T_{eff}$ and $\log g$: $3500 \, {\rm K}
\le T \le 50000 \, {\rm K}$, \& $0 \le \log g \le 5$.  However, as shown
in Figure~\ref{fig:tg}, some of the $\log T_{eff}$ and $\log g$ values
from the stellar tracks are located outside this range, 
in particular for small $\log T_{eff}$ and $\log g$.  For this
reason, empirical M giant spectra (e.g., Bessell et al. 1989, 1991;
Fluks et al. 1994) are often adopted.  
However, as noted by Girardi et al. (2002), the magnitudes from the
ATLAS9 models, and those from the supplementary spectra, do not nicely match at
their boundaries.  We find that the isochrones in this overlap region are
sensitive on how one applies the interpolation between the two libraries.  
The goal of the present paper is simply to present similarities/disimilarities
of the isochrones for different filters, not the most complete isochrones,
so we choose not to supplement the ATALS9 models.  
Consequently, our isochrones do not include M giants cooler
than $T_{eff} \sim 3500$~K, relatively rare metal-rich stars that 
are usually in the thermally pulsing AGB phase of evolution, and are found 
at the upper tip of the red giant branch. 

We similarly choose not to supplement the ATLAS9 models with synthetic spectra
of M, L, \& T dwarfs (low $T_{eff}$ and high $g$), as was done for the Padova
models by Girardi et al. (2002) using the spectra calculated by Allard et al.
(2000).

%%%%%%%%%%%%%%%%%%%%%%%%%%%%%%%%%%%%%%%%%%%%%%%%%%%%%%%%%%%%%%%%%%%%%%%%%%%%%%%%
\section{Isochrones}
\label{sec:isochrones}

We first calculate a table of magnitudes for all spectra in ATLAS9 models in
the $H$ and $K$ band filters, covering a large range in $T_{eff}$,
$\log g$, and [M/H], using equations~(\ref{Mag2}) and (\ref{Mag2NIC}).
We use this table as a set of interpolates for the $T_{eff}$, $\log g$, and
$Z$ values predicted by the stellar evolution models for a given age in order
to estimate synthetic isochrones.

\begin{figure}
%Fig 3
\epsscale{1.0}
\plotone{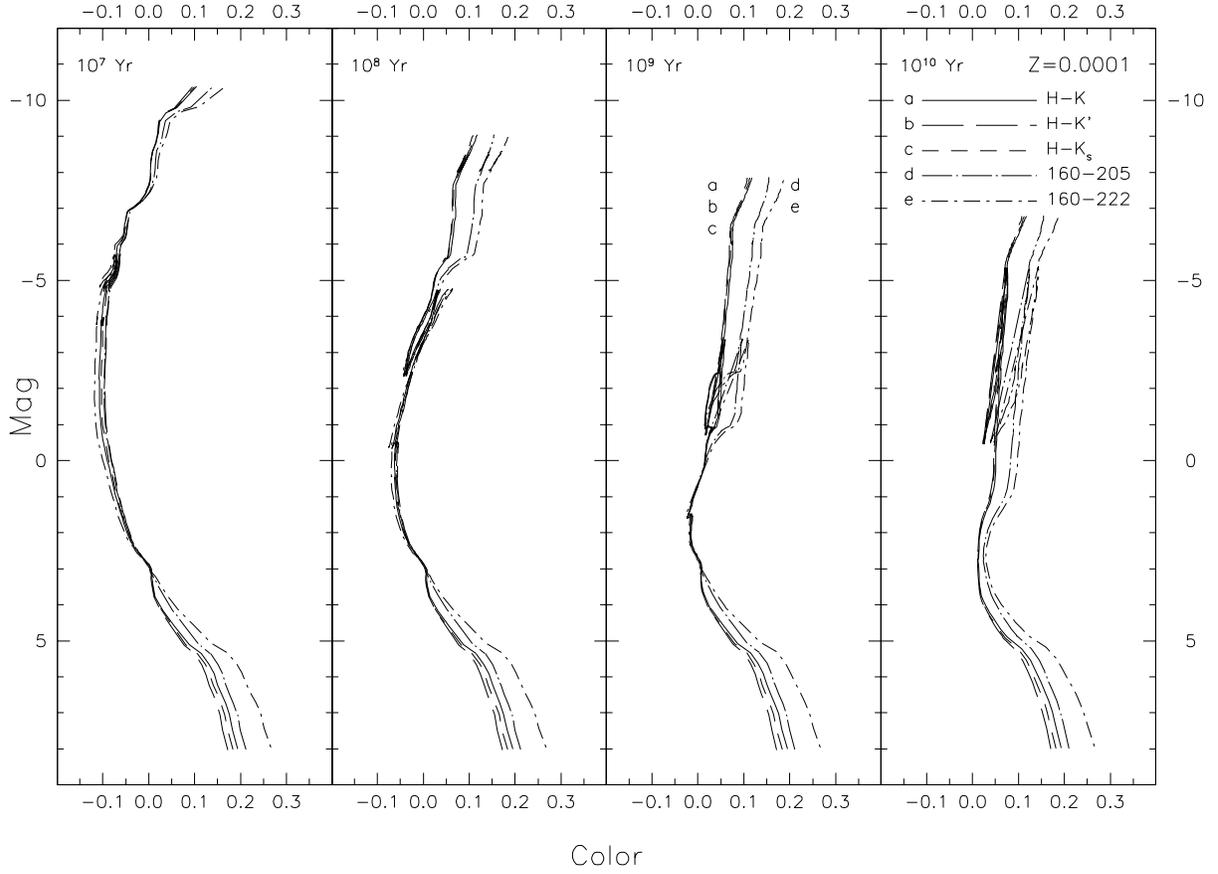}
\caption
{\label{fig:iso1}Isochrones of $Z=0.0001$ model for $H-K$ vs. $K$ (solid),
$H-K'$ vs. $K'$ (long dash), $H-K_s$ vs. $K_s$ (short dash), F160W$-$F205W
vs. F205W (long dash-dot), and F160W$-$F222M vs. F222M (short dash-dot),
in Vega magnitude system.  The three atmospheric filters nearly coincide
at the bright end.  Only the data points
that have $\log T_{eff} \ge 3500$~K and $\log g \ge 0$ are plotted.}
\end{figure}

\begin{figure}
%Fig 4
\epsscale{1.0}
\plotone{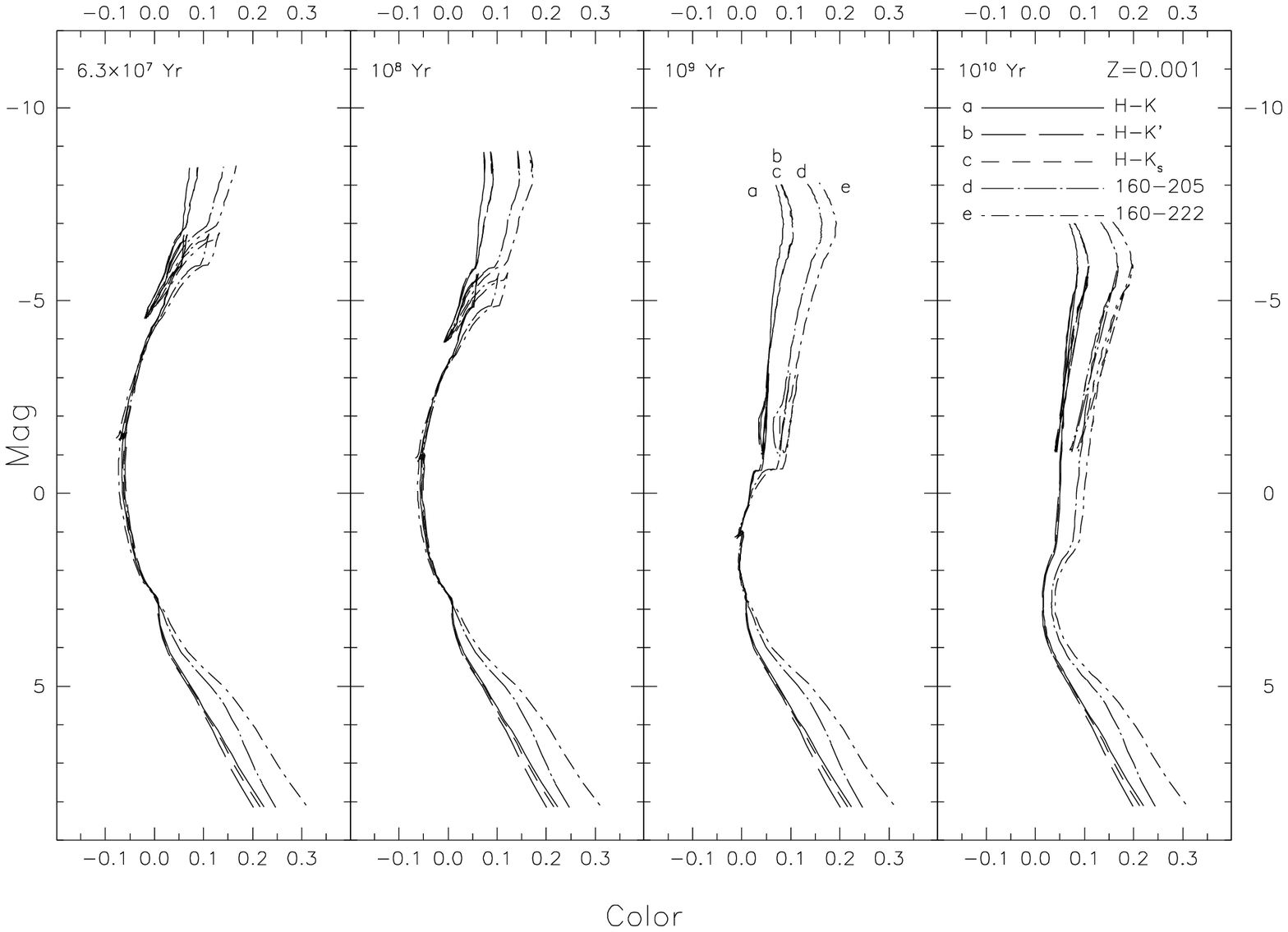}
\caption
{\label{fig:iso2}Same as Figure~\ref{fig:iso1} but for $Z=0.001$ model.
Isochrones for $K'$ and $K_s$ are indistinguishable.}
\end{figure}

\begin{figure}
%Fig 5
\epsscale{1.0}
\plotone{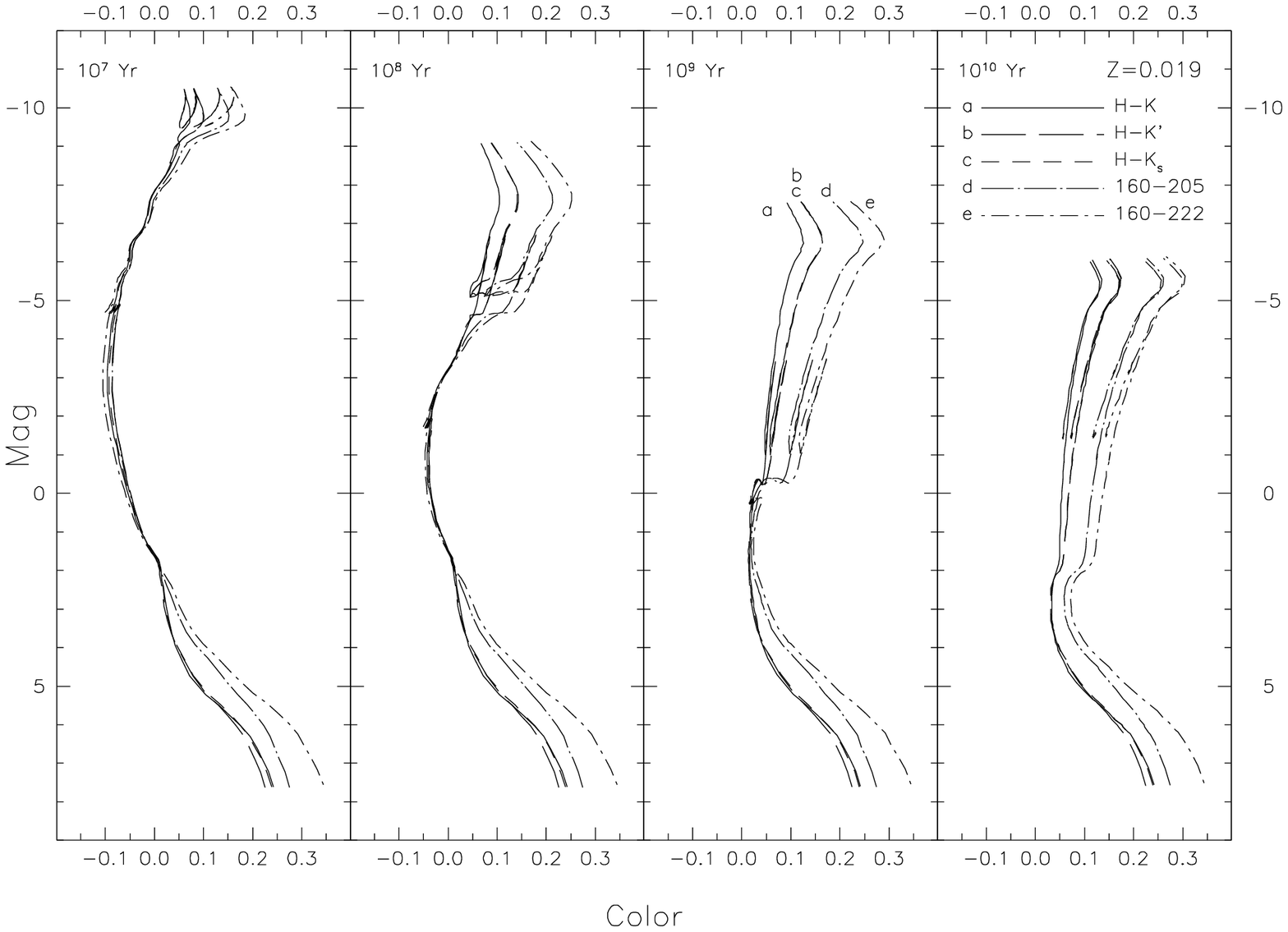}
\caption
{\label{fig:iso3}Same as Figure~\ref{fig:iso1} but for $Z=0.019$ model.
Isochrones for $K'$ and $K_s$ are indistinguishable.}
\end{figure}

\begin{figure}
%Fig 6
\epsscale{1.0}
\plotone{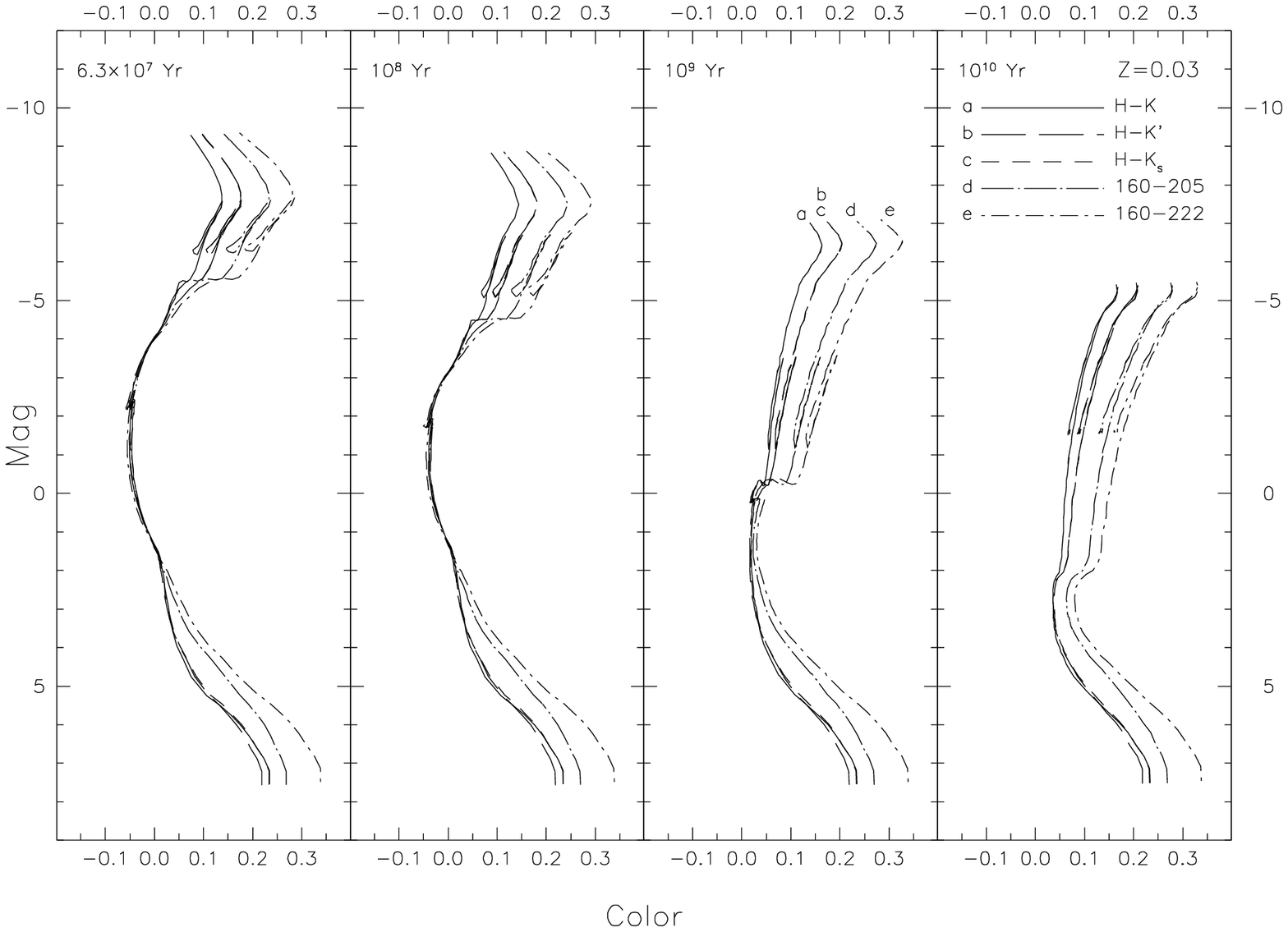}
\caption
{\label{fig:iso4}Same as Figure~\ref{fig:iso1} but for $Z=0.03$ model.
Isochrones for $K'$ and $K_s$ are indistinguishable.}
\end{figure}

Isochrones for $A_\lambda = 0$, calculated in this way, are shown in
Figures~\ref{fig:iso1}$-$\ref{fig:iso4} for four different metallicities,
and four ages.  As expected for color-magnitude diagrams (CMD) in the
infrared, the isochrones have relatively narrow color spreads.  The color
differences between filters is more prominent for the highest metallicity
isochrones.  In most cases, isochrones for $K'$ and $K_s$ are nearly
indistinguishable, and those for F205W and F222M are quite close to each
other.  The largest color difference, 0.18 mag, is seen between the bright
end of the $K$ and F222M isochrones for $Z$ = 0.019 and age = $10^{10}$~yr.
In general, intrinsic color differences of red giants between the
atmospheric filters and the space filters are $\sim$0.1-0.15~mag.

\begin{figure}
%Fig 7
\epsscale{1.0}
\plotone{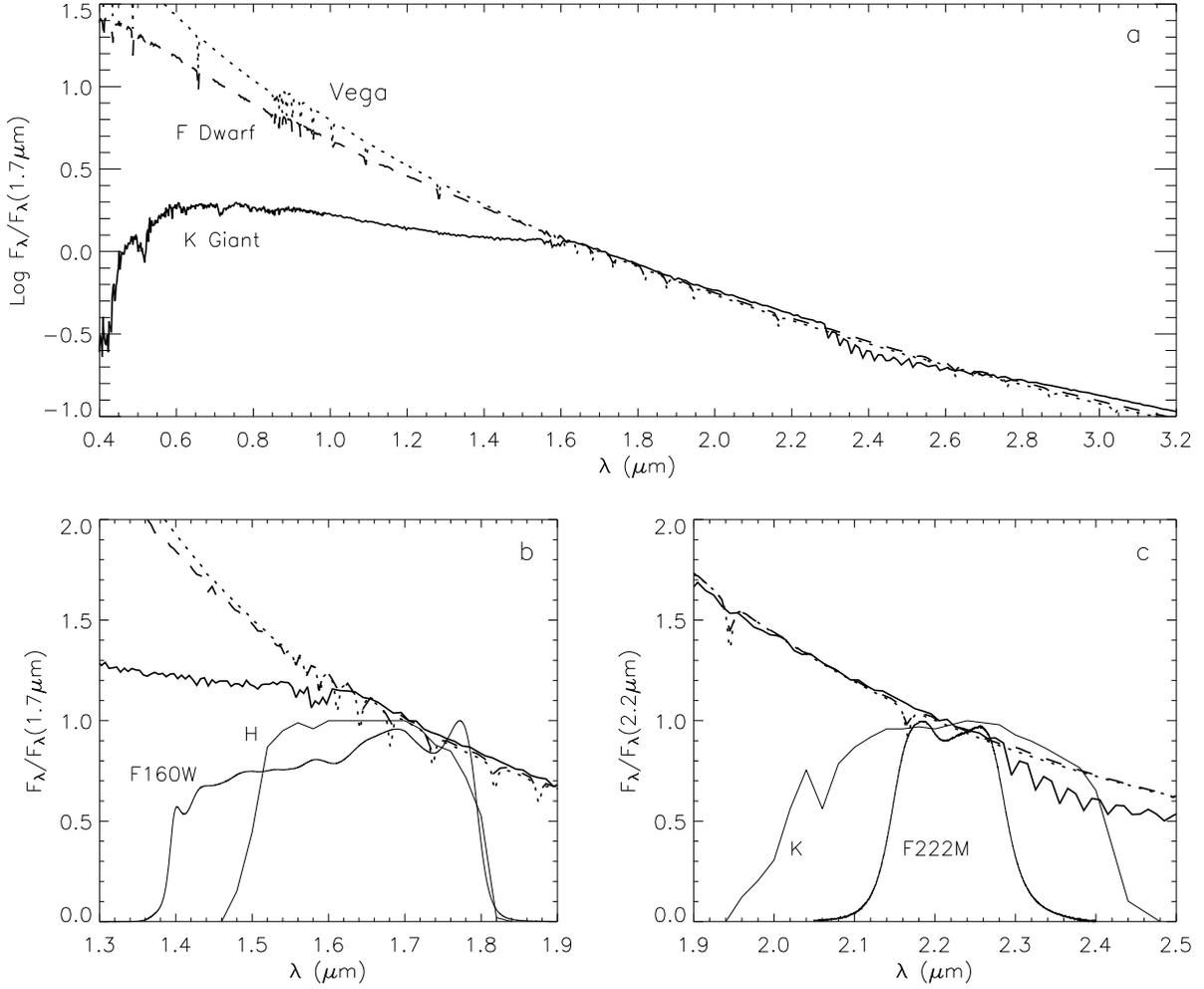}
\caption
{\label{fig:spec}
ATLAS9 stellar spectra for a K giant (thick solid lines;
$T_{eff}=4000~K$, $\log g = 1.0$), an F dwarf (dashed lines; $T_{eff}=7000~K$,
$\log g = 3.5$), and Vega (dotted lines).  Also plotted in panels $b$ and $c$
are the transmission functions ($S_\lambda$) of $H$, $K$, F160W, and F222M
filters (thin solid lines).
}
\end{figure}

Such color differences at the bright end of red giant isochrones between
different filters are due to a kink in the continuum near $1.6 \,
{\rm \mu m}$ and the CO absorption band longward of $2.3 \, {\rm \mu m}$.
Figure~\ref{fig:spec} compares the spectrum of a K giant with
$Z=0.019$ \& Age~=~$10^9$~yr ($K \simeq -5.3$~mag, $T_{eff}=4000~K$,
$\log g = 1.0$), whose F160W$-$F222M color is 0.13~mag redder than
its $H-K$ color, and that of an F dwarf ($K \simeq 0.0$~mag,
$T_{eff}=7000~K$, $\log g = 3.5$), whose F160W$-$F222M color is
only 0.01~mag redder than its $H-K$ color.  The spectrum of an F dwarf
is very close to that of Vega, and has no strong features in $H$
and $K$ bands, making all colors in our filter sets
nearly zero.  On the other hand,
the spectrum of a K giant 1) significantly deviates from that of Vega
at the wavelengths shortward of $1.6 \, {\rm \mu m}$ due to the H$^-$
opacity in cool stars (J. Valenti 2005, private communication), which
causes F160W magnitudes to be fainter than $H$ magnitudes, and 2) has
the CO absorption band longward of $2.3 \, {\rm \mu m}$, which causes
$K$ magnitudes to be fainter than F222M magnitudes.  These two effects
make F160W$-$F222M redder than $H-K$, and the same analysis
can be applied to other filter sets.

\begin{figure}
%Fig 8
\epsscale{1.0}
\plotone{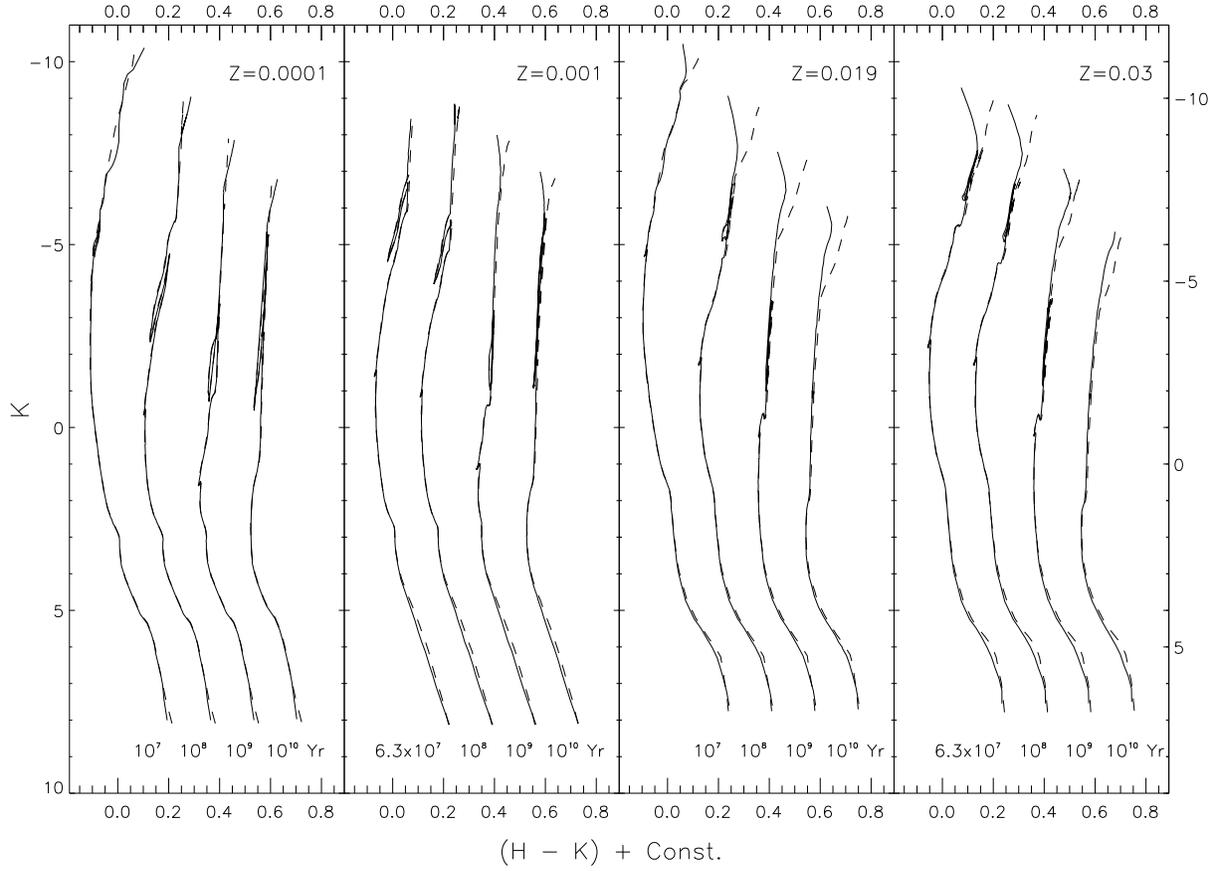}
\caption
{\label{fig:padova}
$H-K$ vs. $K$ isochrones calculated in the present study (solid lines)
and those by Girardi et al. (2002; dashed lines).  Only the data points
that have $\log T_{eff} \ge 3500$~K and $\log g \ge 0$ are plotted.
}
\end{figure}

As an independent check of our procedure, we compare our $H-K$ vs. $K$
isochrones to those calculated by Girardi et al. (2002) in
Figure~\ref{fig:padova}.  The isochrones match nicely, except
at the extremes.  The discrepancy in the bright end
is caused by the empirical M giant spectra that Girardi et al. (2002)
added to their spectral library, and that in the faint end is by the
late M dwarf spectra.  The discrepancies are considerable only at
the top 1 or 2 magnitudes of the isochrone, where only a small
number fraction of giants reside.

\begin{figure}
%Fig 9
\epsscale{0.8}
\plotone{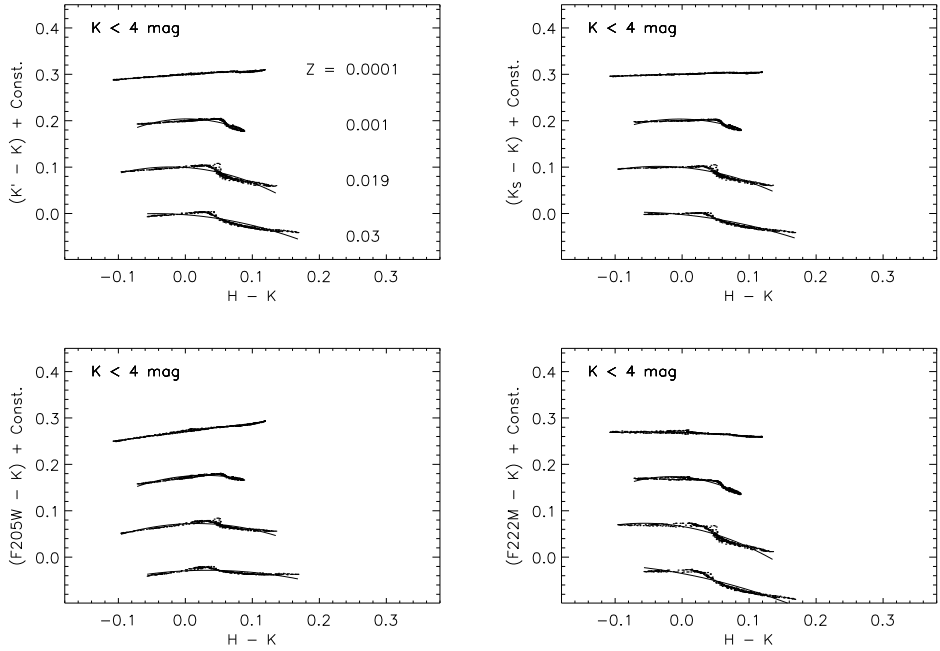}
\caption
{\label{fig:trans1a}Magnitude differences between $K$ and the other $K$-band
filters as a function of $H-K$ color for all ages considered in the present
paper ($10^7$, $10^8$, $10^9$, \& $10^{10}$ yr for $Z=0.0001$ \& 0.019 models,
and $6.3\times 10^7$, $10^8$, $10^9$, \& $10^{10}$ yr for $Z=0.001$
\& 0.03 models), for $K$-band magnitudes brigher than 4~mag.  Best-fit
2nd-order polynomials are also shown.  Only the data points that have
$\log T_{eff} \ge 3500$~K and $\log g \ge 0$ are plotted.}
\end{figure}

\begin{figure}
%Fig 10
\epsscale{0.8}
\plotone{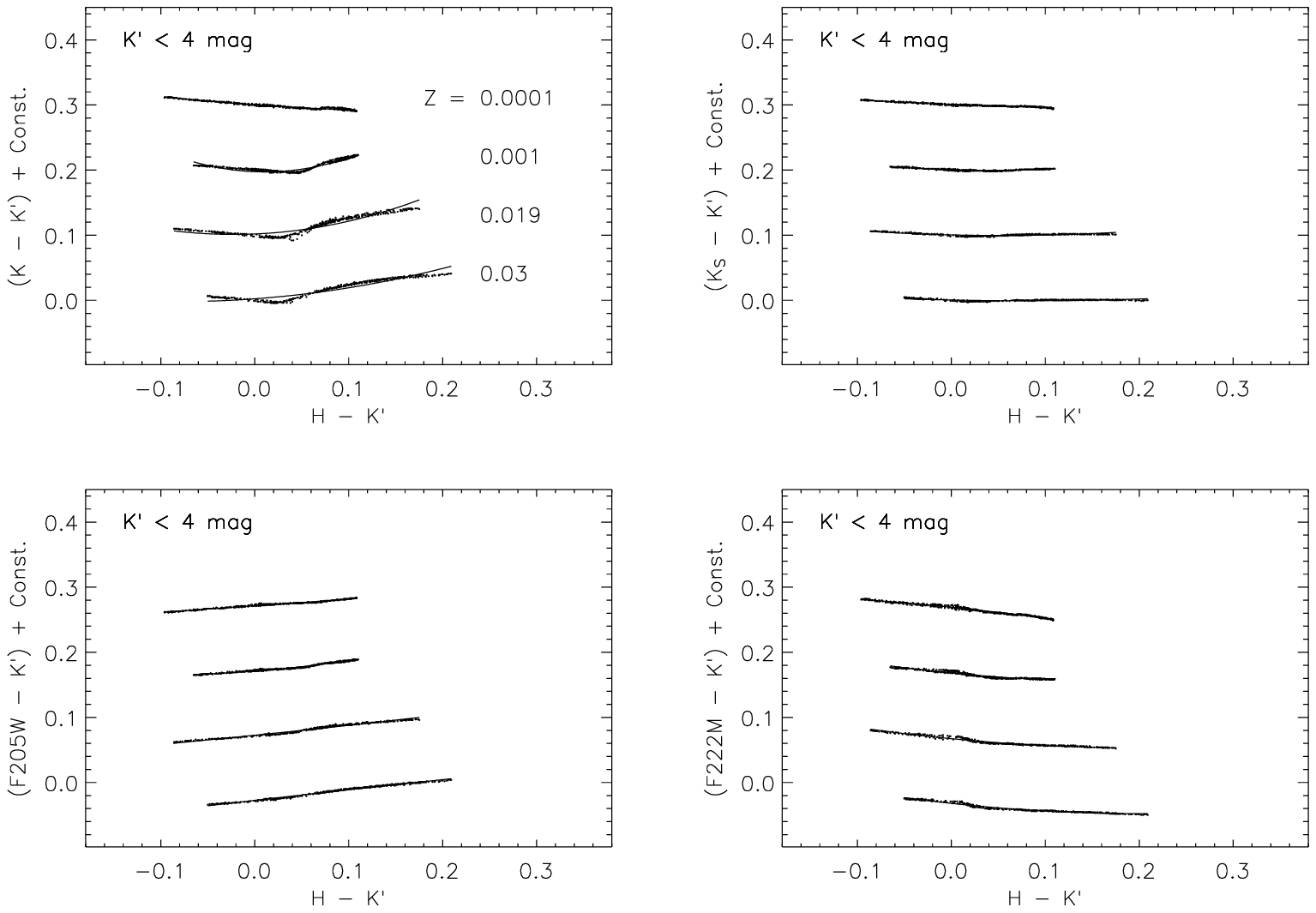}
\caption
{\label{fig:trans1b}Same as Figure~\ref{fig:trans1a} but for magnitude
differences between $K'$ and the other $K$-band filters as a function of
$H-K'$ color.}
\end{figure}

\begin{figure}
%Fig 11
\epsscale{0.8}
\plotone{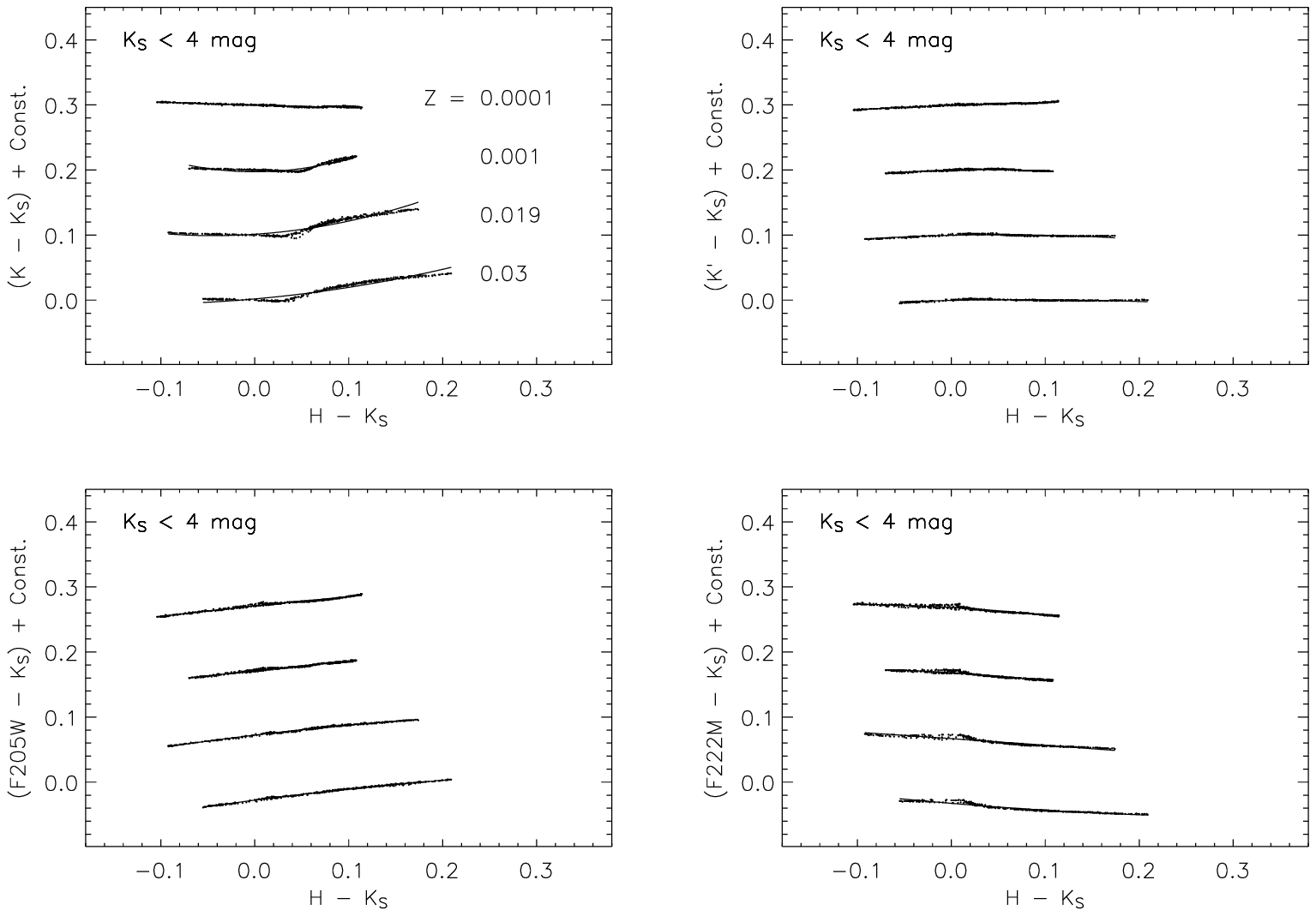}
\caption
{\label{fig:trans1c}Same as Figure~\ref{fig:trans1a} but for magnitude
differences between $K_s$ and the other $K$-band filters as a function of
$H-K_s$ color.}
\end{figure}

\begin{figure}
%Fig 12
\epsscale{0.8}
\plotone{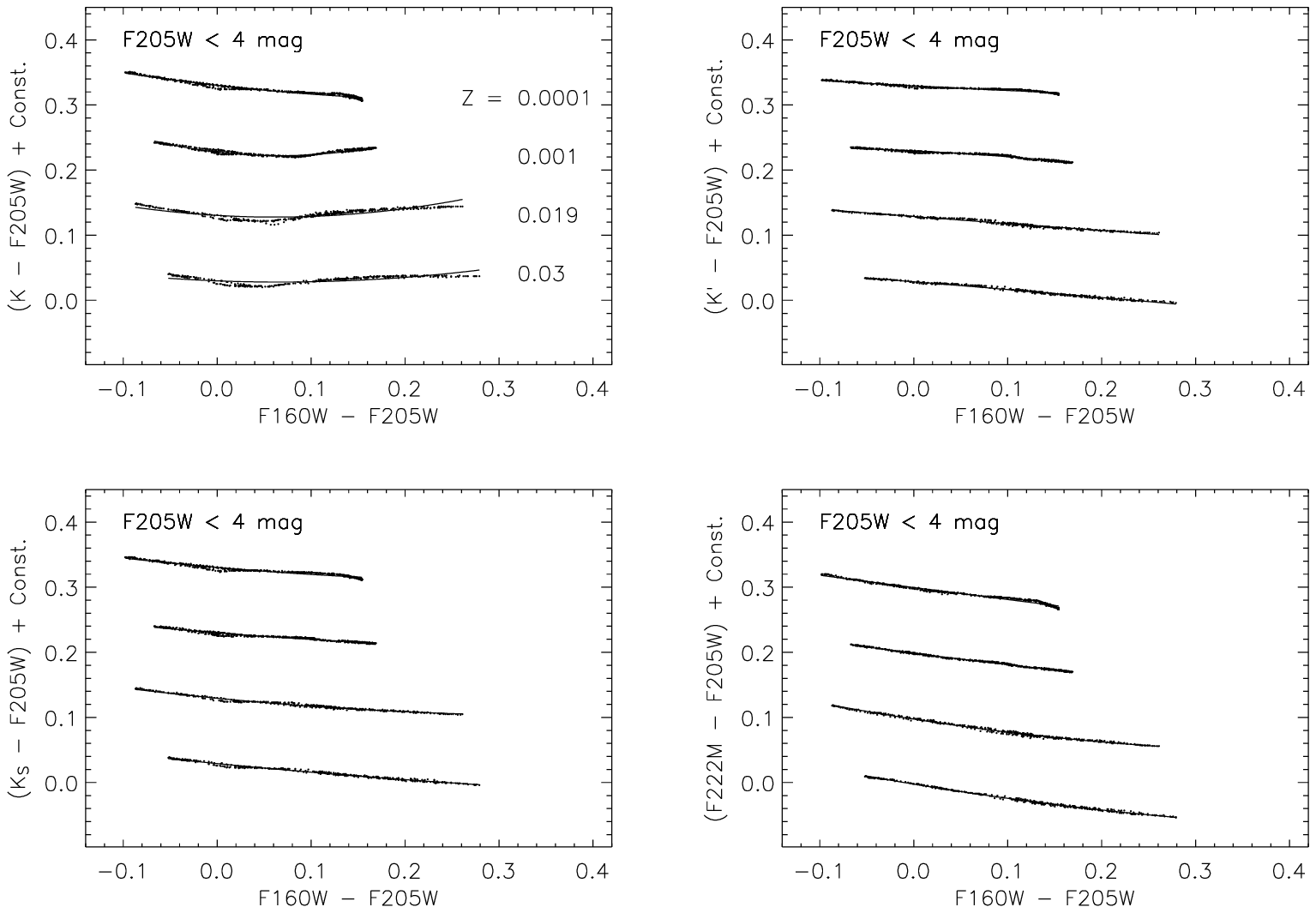}
\caption
{\label{fig:trans1d}Same as Figure~\ref{fig:trans1a} but for magnitude
differences between F205W and the other $K$-band filters as a function of
F160W$-$F205W color.}
\end{figure}

\begin{figure}
%Fig 13
\epsscale{0.8}
\plotone{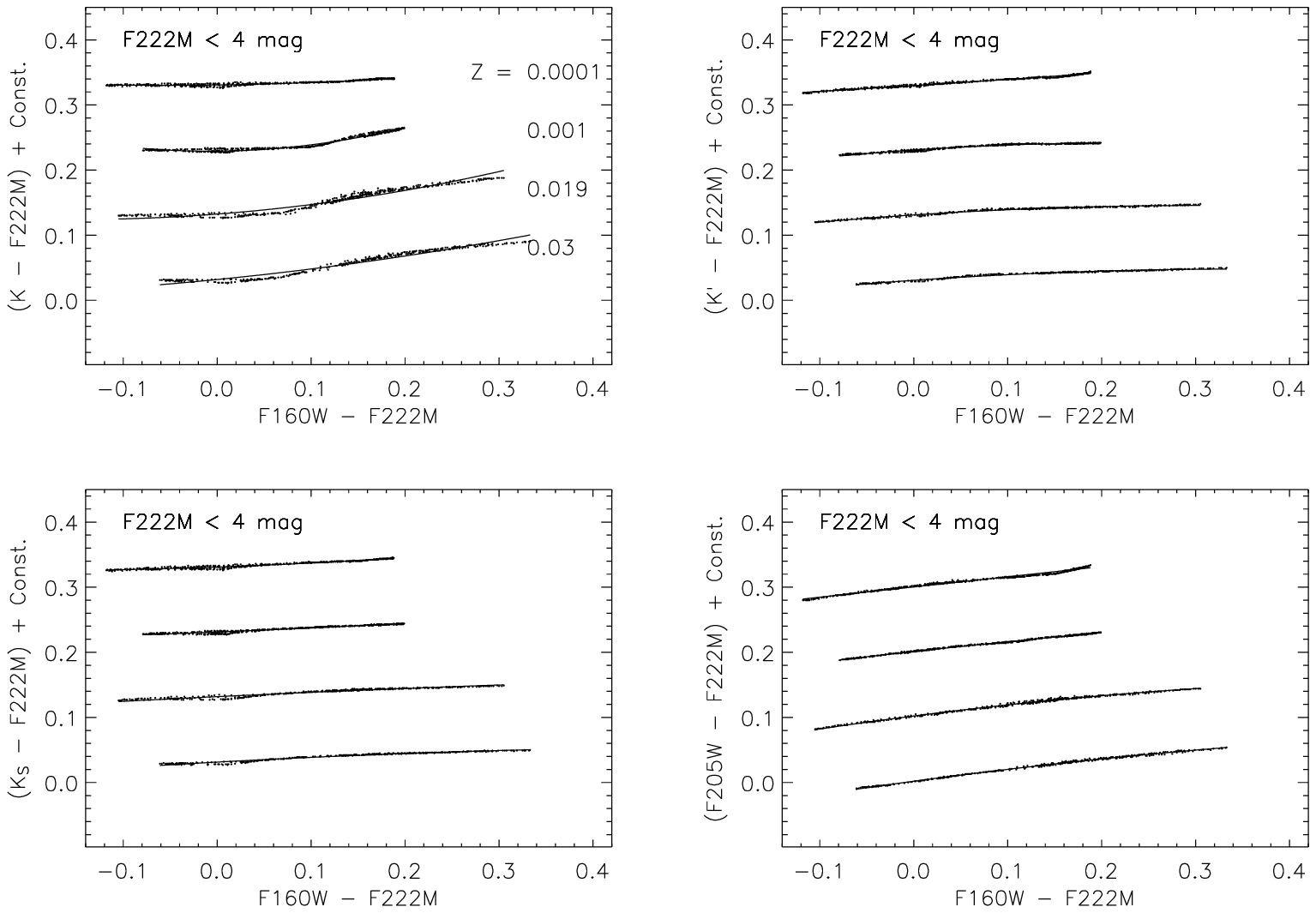}
\caption
{\label{fig:trans1e}Same as Figure~\ref{fig:trans1a} but for magnitude
differences between F222M and the other $K$-band filters as a function of
F160W$-$F222M color.}
\end{figure}

Magnitude transformations between $K$-band filters can be
obtained from our isochrones.  Figures~\ref{fig:trans1a} through
\ref{fig:trans1e} show magnitude differences between $K$-band filters
as a function of color for all ages considered 
($10^7$, $10^8$, $10^9$, \& $10^{10}$ yr for $Z = 0.0001$ \& 0.019 models,
and $6.3\times 10^7$, $10^8$, $10^9$, \& $10^{10}$ yr for
$Z = 0.001$ \& 0.03 models), for $K$-band magnitudes brighter than
4~mag.  We find that the magnitude difference can be well fit
by a quadratic function for $M < 4$~mag, and by a separate straight line
for $M > 4$~mag (we do not present figures for $M > 4$~mag).
The largest residuals from the fit are 0.019~mag for the former and 0.009~mag
for the latter.  The coefficients of the best-fit functions are
presented in Tables~\ref{table:trans1} and \ref{table:trans2}, along with
the residuals and fitting ranges.

\begin{figure}
%Fig 14
\epsscale{1.0}
\plotone{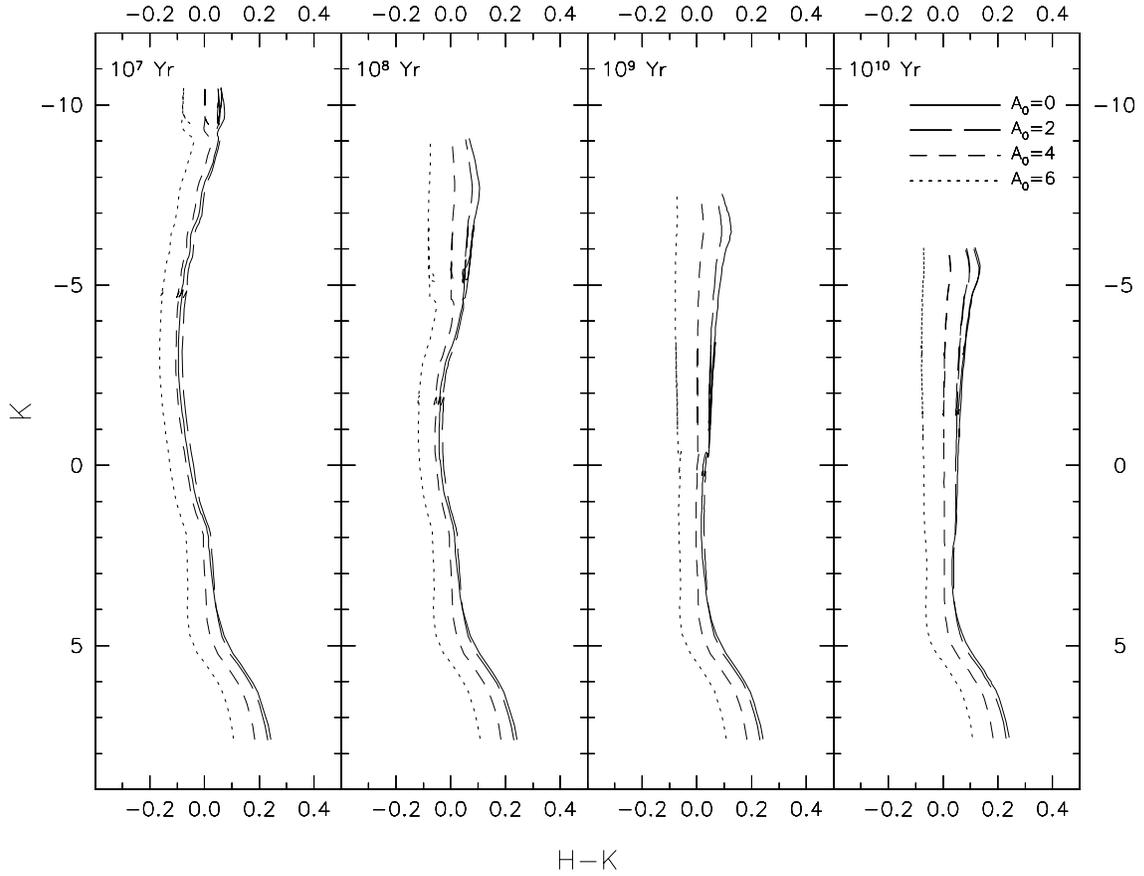}
\caption
{\label{fig:red1}Dereddened $H-K$ vs. $K$ isochrones of $Z=0.019$ model
with $A_0$=0 (solid), $A_0$=2 (long dash), $A_0$=4 (short dash), and $A_0$=6
(dots).  The isochrones are dereddened by an amount of $A_0 (\lambda_c /
\lambda_0)^{-1.61}$.  Only the data points that have $\log T_{eff} \ge
3500$~K and $\log g \ge 0$ are plotted.}
\end{figure}

\begin{figure}
%Fig 15
\epsscale{1.0}
\plotone{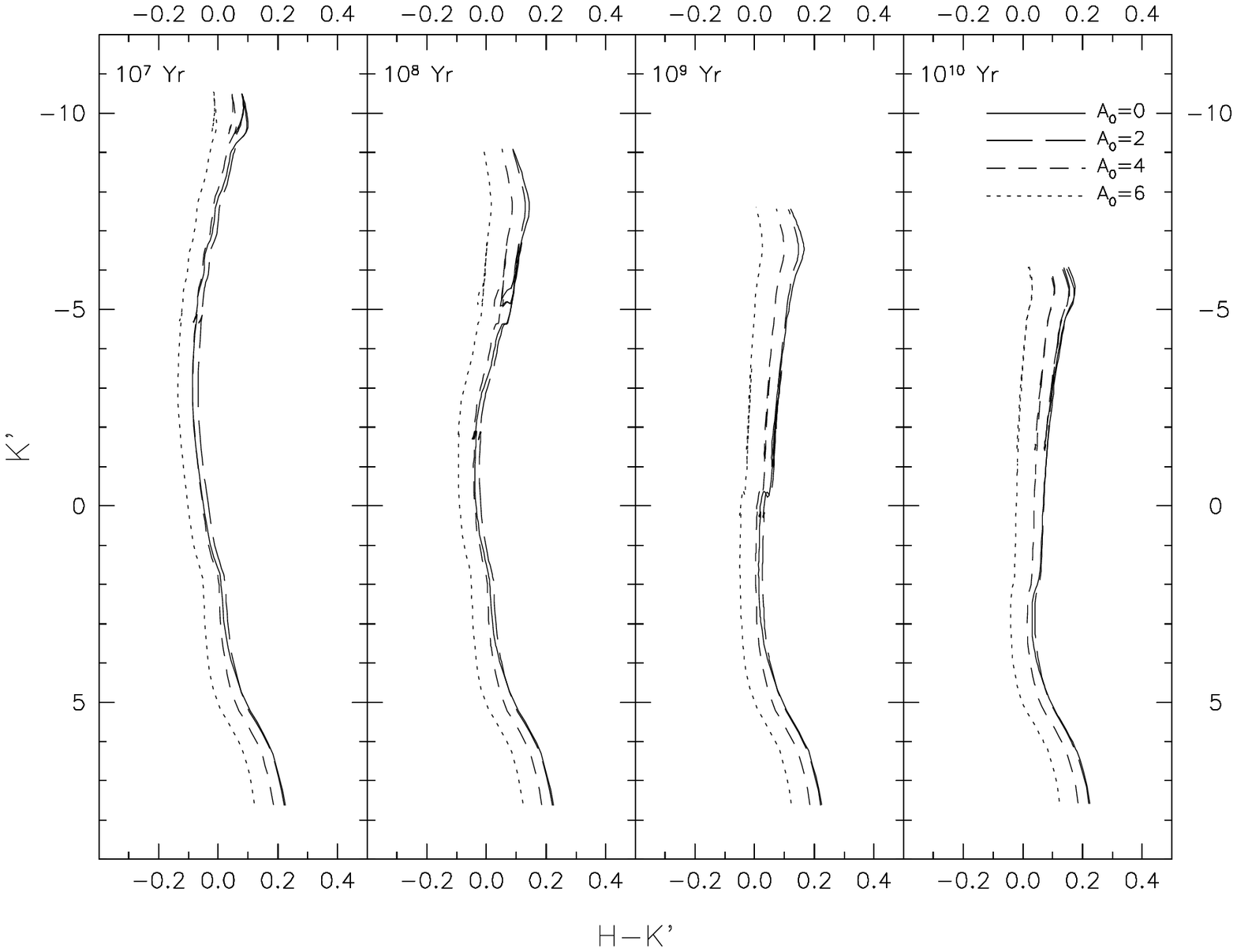}
\caption
{\label{fig:red2}Same as Figure~\ref{fig:red1} but for $H-K'$ vs. $K'$.}
\end{figure}

\begin{figure}
%Fig 16
\epsscale{1.0}
\plotone{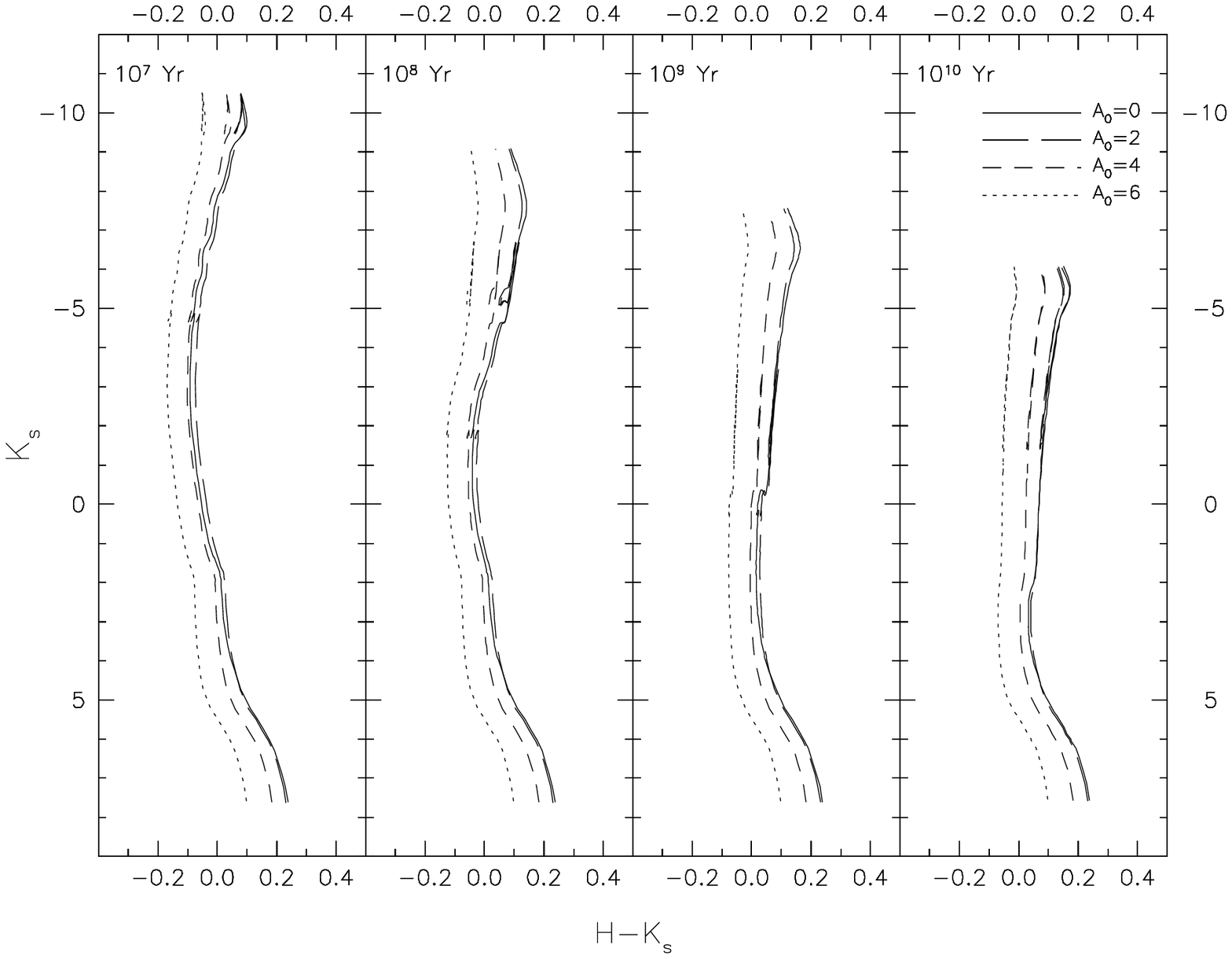}
\caption
{\label{fig:red3}Same as Figure~\ref{fig:red1} but for $H-K_s$ vs. $K_s$.}
\end{figure}

\begin{figure}
%Fig 17
\epsscale{1.0}
\plotone{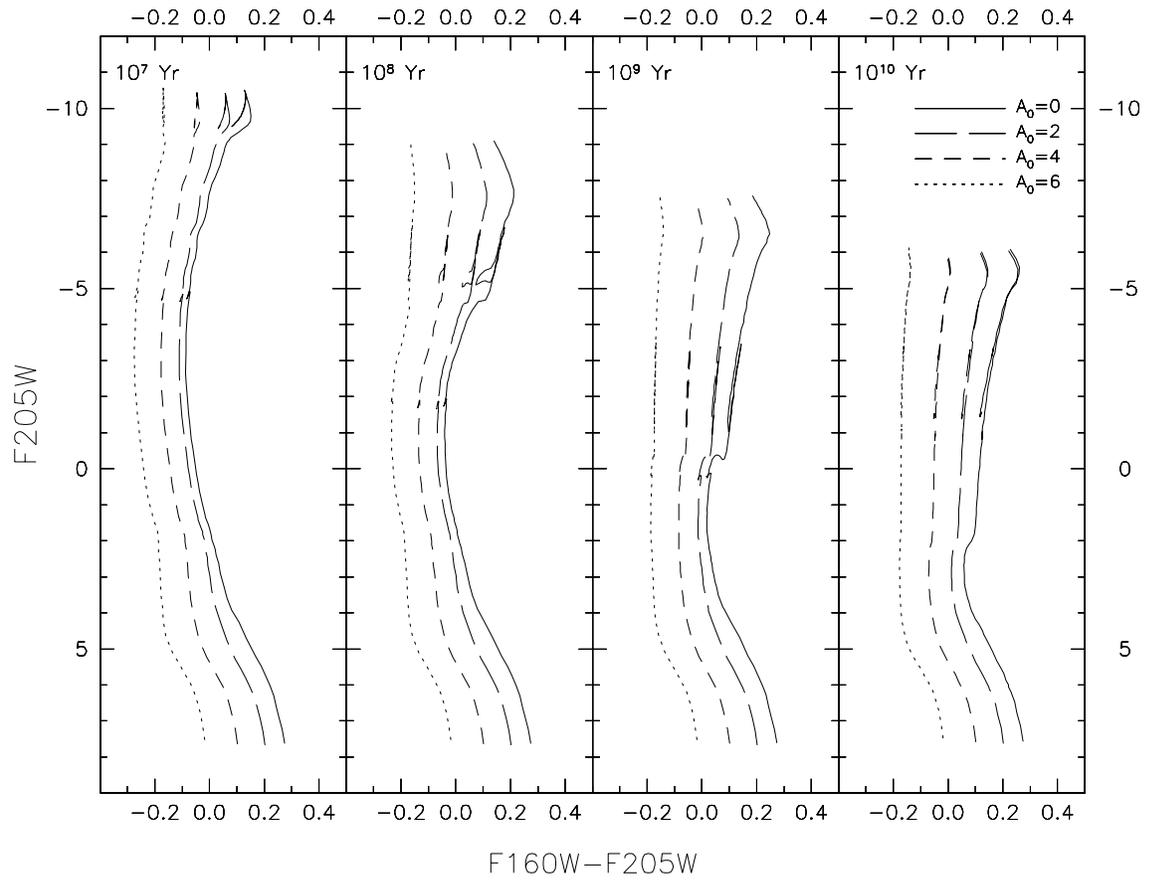}
\caption
{\label{fig:red4}Same as Figure~\ref{fig:red1} but for F160W$-$F205W vs. F205W.}
\end{figure}

\begin{figure}
%Fig 18
\epsscale{1.0}
\plotone{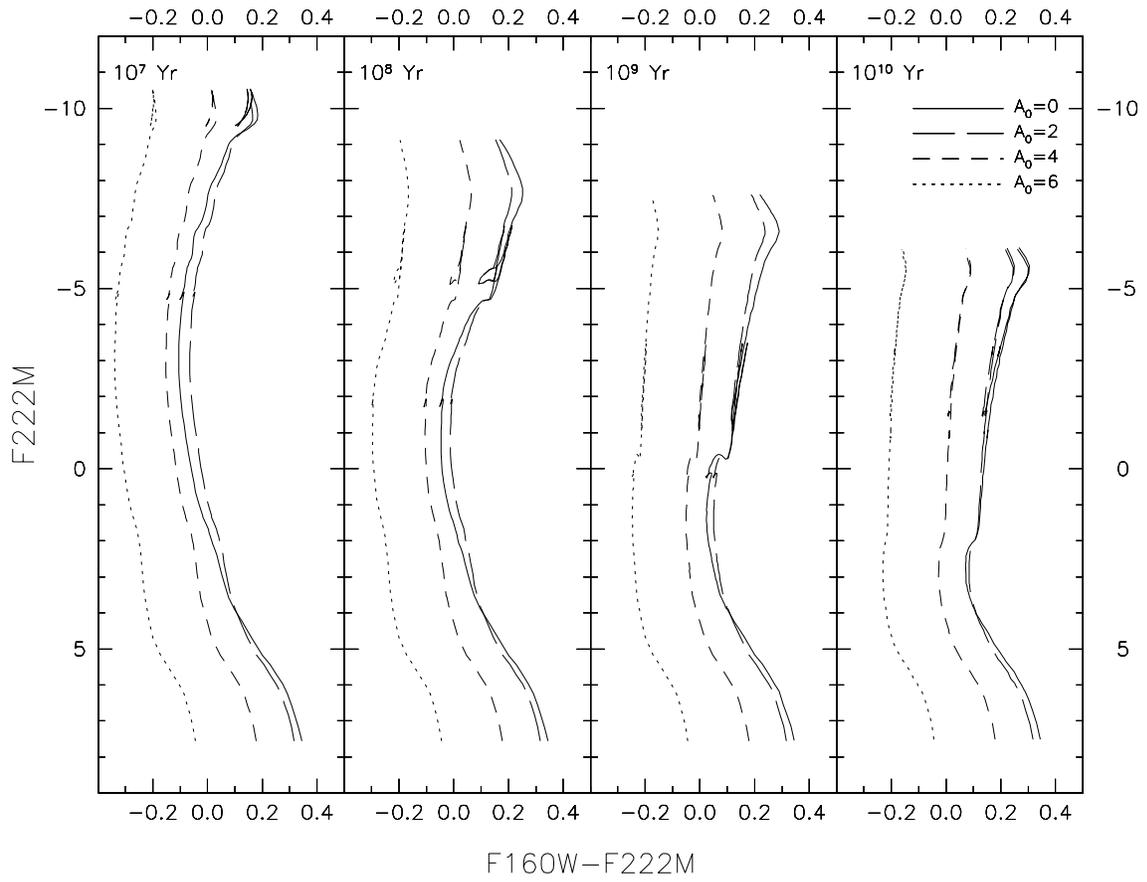}
\caption
{\label{fig:red5}Same as Figure~\ref{fig:red1} but for F160W$-$F222M vs. F222M.}
\end{figure}

Extinction in the $K$-band is much lower than that in visual
bands, but there are still cases where the $K$-band extinction is
significant, such as the central region of the Milky Way ($A_K\sim3$).
Here, we calculate isochrones with $K$-band extinctions up to 6 mag,
some of which are shown in Figures~\ref{fig:red1}$-$\ref{fig:red5}.
For $A_\lambda$ in equations~(\ref{Mag2}) \& (\ref{Mag2NIC}), we adopt a
power-law extinction law for wavelengths between $H$ and $K$ bands:
\begin{equation}
\label{extinction}
	A_\lambda = A_0 \left ( \frac{\lambda}{\lambda_0}
			\right )^{-\alpha},
\end{equation}
where we choose $\lambda_0 = 2.2 \, \mu$m, and $A_0$ is the extinction at
$\lambda_0$.  When assuming that the transmission functions of $H$ and $K$
filters are Dirac delta functions centered at 1.65~$\mu$m and 2.2~$\mu$m,
respectively, the extinction law by Rieke, Rieke, \& Paul (1989) gives
$\alpha=1.55$.
However, as discussed later in this section, the apparent extinction behaviour
of isochrones in the CMD can be different from the actual extinction law,
due to non-zero width and asymmetry of the filter transmission functions.
We find that $\alpha=1.61$ makes the isochrone for the $Z = 0.019$,
Age = $10^9$~yr model behave in the CMD as if it follows an extinction law
with $\alpha=1.55$.  We choose this particular isochrone for calibrating
the extinction law, with an assumption that the stars used in Rieke et al.
(1989) to derive their extinction law, which are the stars in the central
parsec of our Galaxy, can be represented by such metallicity and age.

For the sake of easier comparison, isochrones in
Figures~\ref{fig:red1}$-$\ref{fig:red5} are dereddened by an amount
of $A_0 ( \lambda_c / \lambda_0 )^{-1.61}$, where the central wavelength
of the filter $\lambda_c$ is defined by
\begin{equation}
\label{lambda_c}
	\lambda_c = \frac{\int_{\lambda 1}^{\lambda 2} S_\lambda \lambda
			  d \lambda}
			 {\int_{\lambda 1}^{\lambda 2} S_\lambda d \lambda},
\end{equation}
and is given in Table~\ref{table:lambdac}.  As seen in the figures,
isochrones with larger extinction values are more vertically straight,
in particular for $10^9$ and $10^{10}$~yr models.  This is probably
because the extinction makes the longer wavelength part of the spectrum
more important, which is closer to the Rayleigh-Jeans regime of the spectrum.

Since we have dereddened the isochrones with the known amount of
extinction at $\lambda_c$, all the dereddened isochrones with different
extinction values in Figures~\ref{fig:red1}$-$\ref{fig:red5} should
be coincident, if the filter transmission functions were
Dirac delta functions centered at $\lambda_c$.
The figures show that the dereddened isochrones
do not align, and the amount of misalignment is rather large for some
filters.  This implies that the inferred amount of extinction can be
sensitively dependent on the shape of the filter transmission function.

\begin{figure}
%Fig 19
\epsscale{1.0}
\plotone{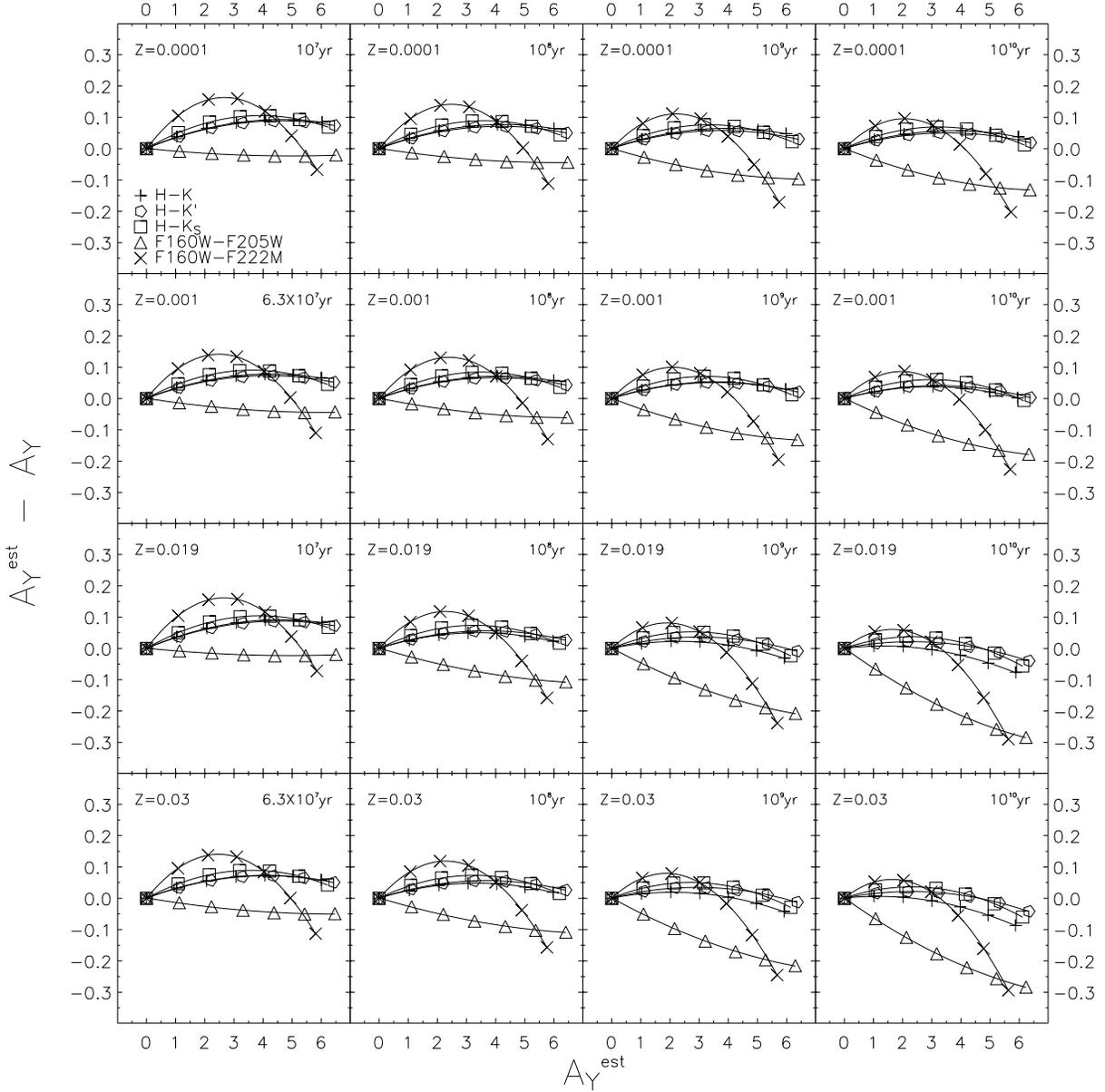}
\caption
{\label{fig:adiff1}The difference between the extinction values
that are estimated by equation~(\ref{A_est}) with the colors from our
reddened isochrones and the actual extinction values.  A constant
value of 1.55 is used for $\alpha$ in equation~(\ref{A_est}).
The extinction of each isochrone has been estimated with the mean color
(for $A^{est}_Y$) and the mean magnitude (for $A_Y$) of the reddened isochrone
data points whose intrinsic $K$-band magnitudes are between $-$6 and 0 mag.
}
\end{figure}

Let us see how this will affect the analyses of actual observations.
When one estimates the amount of extinction from an observed near-infrared
CMD, where isochrones are nearly vertically
straight, one converts an observed color excess to an extinction value
following an assumed extinction law, which usually has the form of a power-law.
When one has photometric data from a set of two filters, $X$ and $Y$, the
amount of extinction can be estimated by
\begin{eqnarray}
\label{A_est}
	A_Y^{est} & = & \frac{(m_X-m_Y)-(m_X-m_Y)_0}{A_X/A_Y-1} \cr
                  & = & \frac{(m_X-m_Y)-(m_X-m_Y)_0}
			     {(\lambda_X/\lambda_Y)^{-\alpha}-1},
\end{eqnarray}
where $m_X$ \& $m_Y$ and $\lambda_X$ \& $\lambda_Y$ are the magnitudes and
the central wavelengths of the two filters, respectively, and subscript 0
denotes the intrinsic value.  For estimating extinction from our isochrones,
we first use $\alpha=1.55$.  Figure~\ref{fig:adiff1}
shows the difference between the inferred extinction values,
using equation~(\ref{A_est}) and colors from our
reddened isochrones, and the actual extinction values.  Here, the extinction
of each isochrone has been calculated using the mean color (for $A^{est}_Y$)
and magnitude (for $A_Y$) of the reddened isochrone data points
having $K$-band magnitudes between $-$6 and 0 mag.  As the figure shows,
the differences between
estimated and actual extinction values are larger for the space filter sets
in general.  The largest relative difference is $\sim 10$~\%, and the largest
absolute difference is 0.29 mag.  Note that the extinction estimates
for the $Z = 0.019$ and Age = $10^9$~yr model inferred from $H$ \& $K$ are very
close to the actual extinction values, justifying our choice of
$\alpha=1.61$ for equation~(\ref{extinction}).

\begin{figure}
%Fig 20
\epsscale{0.6}
\plotone{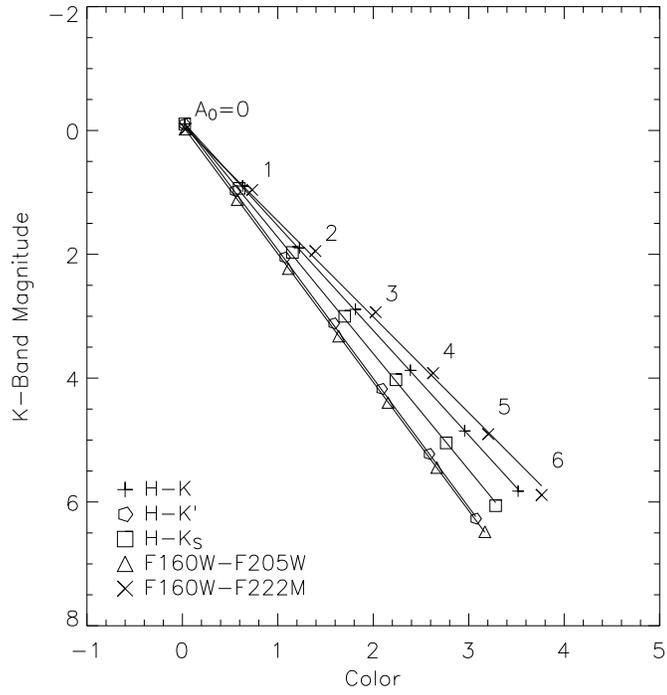}
\caption
{\label{fig:extlaw}Reddened magnitudes of $K$-band filters and reddened
colors for the $Z=0.019$ \& Age = $10^9$~yr isochrone data point whose
intrinsic $K$ magnitude is 0.  Also shown are the best-fit straight
lines that go through the data point for $A_\lambda = 0$ for each filter set.
}
\end{figure}

\begin{figure}
%Fig 21
\epsscale{1.0}
\plotone{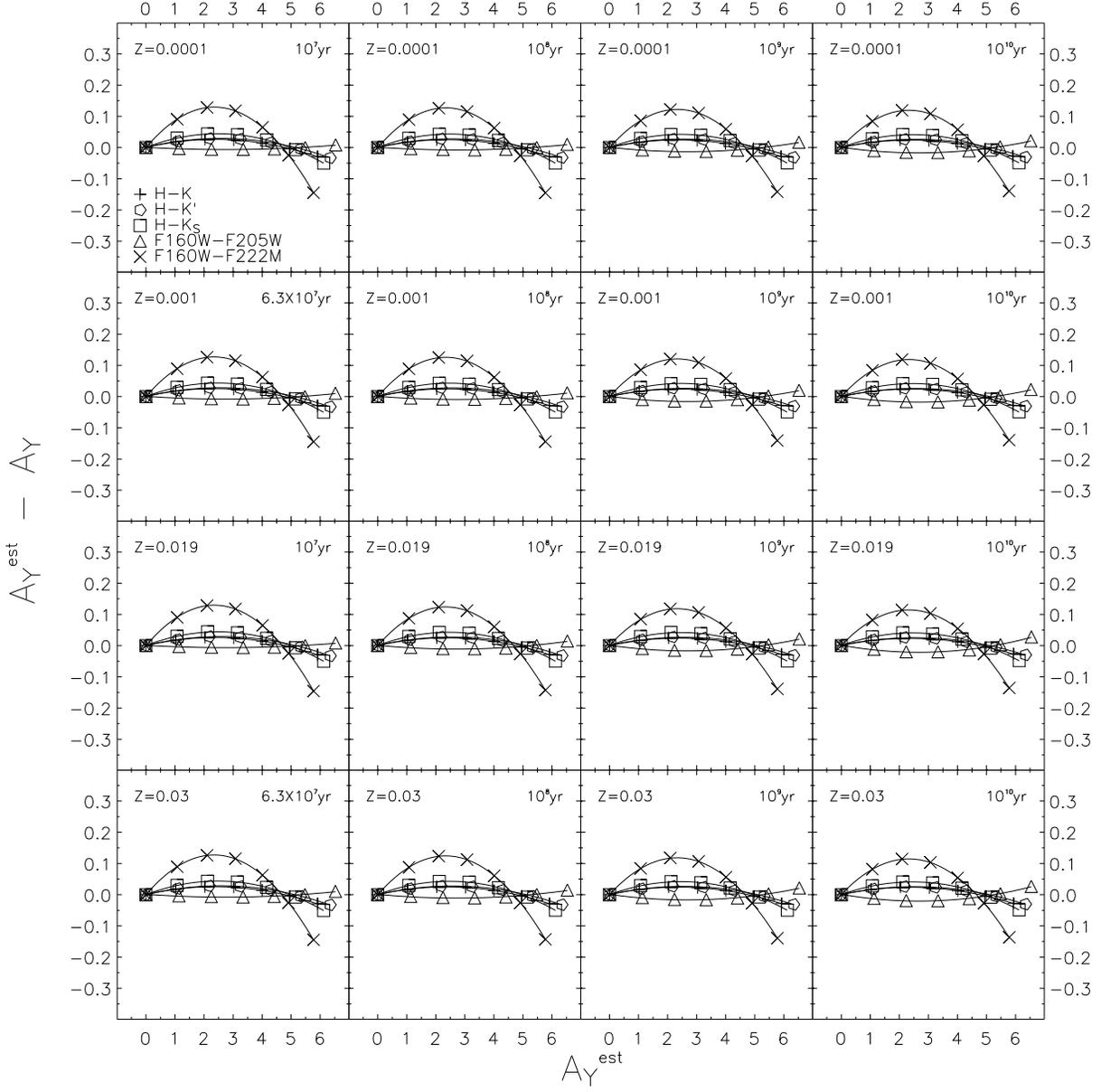}
\caption
{\label{fig:adiff3}Same as Figure~\ref{fig:adiff1} but using $\alpha_{eff}$
for equation~(\ref{A_est}).
}
\end{figure}

The problems seen in Figure~\ref{fig:adiff1} can be alleviated by finding
an ``effective extinction slope'' for each filter set and isochrone model,
which is defined such that it better describes the extinction behaviour
in the CMD.  Figure~\ref{fig:extlaw} shows reddened $K$-band magnitudes and 
colors for the $Z = 0.019$ \& Age = $10^9$~yr isochrone
(the figure only shows an isochrone data point whose intrinsic $K$
magnitude is 0, as an example).  The reddened magnitudes and colors of
each filter set form a nearly straight line (except for filter set
F160W \& F222M), thus we may assume that $A_X/A_Y$ in equation~(\ref{A_est})
is not a function of extinction.  Then the effective slope of the extinction,
$\alpha_{eff}$, can be calculated by
\begin{equation}
\label{alpha_eff}
	\alpha_{eff} = - \frac{ \log (1+1/b) }{ \log (\lambda_X/\lambda_Y) },
\end{equation}
where $b$ is the slope of the straight line that fits the distribution of
reddened magnitudes vs. reddened colors, as in Figure~\ref{fig:extlaw}.
We calculate $b$ for data points of each isochrone whose intrinsic
$K$ magnitudes are between $-6$ and 0~mag, and take an average for
each isochrone model.
When finding the best-fit straight line, we forced the line to include the 
$A_0 = 0$ point, i.e., we fitted $y=y_0+ b(x-x_0)$, instead of $y=a+bx$,
where subscript 0 denotes the values for $A_0 = 0$.
Table~\ref{table:alpha_eff} shows the averages and standard deviations
of $\alpha_{eff}$ values for each isochrone model.
Note that the standard deviations of $\alpha_{eff}$ in an isochrone is
generally much smaller than the differences of average $\alpha_{eff}$ values
between different isochrones.
The average $\alpha_{eff}$ values range from 1.489 to 1.574,
which are 7.5~\% to 2.2~\% smaller than the original $\alpha$ value we
adopted for extinction, 1.61.
As seen in Figure~\ref{fig:adiff3}, extinction values estimated by
equation~(\ref{A_est}) with $\alpha_{eff}$ are now very close to
the actual values, except for filter set F160W \& F222M, whose extinction
behaviour in the CMD is poorly described by a straight line.

\begin{figure}
%Fig 22
\epsscale{1.0}
\plotone{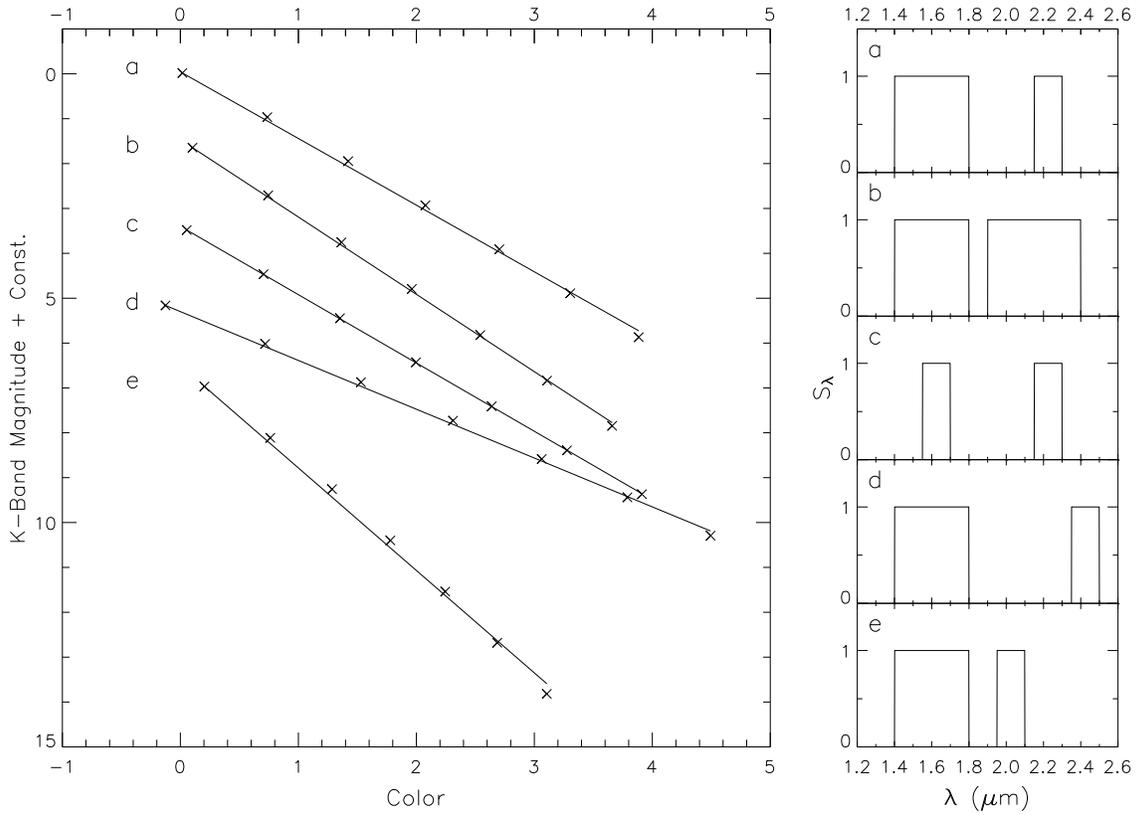}
\caption
{\label{fig:nonlin}Reddened magnitudes and colors (crosses) for the $Z=0.019$
\& Age = $10^9$~yr isochrone data point whose intrinsic $K$ magnitude is 0,
for five imaginary filter sets whose transmission functions are shown in the
right panel.  Also shown are the best-fit straight lines that go through
the data point for $A_\lambda = 0$ for each filter set.
}
\end{figure}

The nonlinear extinction behavior of the filter set F160W \& F222M
appears to be due to a significant width difference in the two filters:
F160W covers $\sim 0.4 {\rm \mu m}$ while F222M covers only $\sim 0.15
{\rm \mu m}$.  Here we discuss the effect of the width difference in terms
of an extinction-weighted central wavelength,
\begin{equation}
\label{lambda_c_eff}
	\lambda_A = \frac{\int_{\lambda 1}^{\lambda 2} 10^{-0.4 A_\lambda}
			  S_\lambda \lambda d \lambda}
			 {\int_{\lambda 1}^{\lambda 2} 10^{-0.4 A_\lambda}
			  S_\lambda d \lambda}.
\end{equation}
As the amount of extinction $A_0$ in equation~(\ref{extinction}) becomes
large, $\lambda_A$ shifts to a longer wavelength because $A_\lambda$
decreases with an increasing wavelength.  However, the amount of this shift
is limited by the filter width.  When the widths of two filters
differ significantly, the relative shifts in $\lambda_A$ 
become considerably different for the two filters.  Since the slope of
the reddening vector in the CMD is determined by a ratio between
$\lambda_A$'s of the two filters (see eq.~\ref{A_est}), the slope of
the reddening vector will become a function of $A_0$ when the ratio
between $\lambda_A$'s is sensitively dependent on $A_0$, i.e. when
the widths of two filters differ significantly.  Figure~\ref{fig:nonlin}
shows the effect of the filter width by comparing the extinction behaviour
of five imaginary filter sets.  Filter set $a$ represents F160W \& F222M,
and its extinction behaviour in the CMD is not linear, as in the real
F160W \& F222M set.  But this nonlinearity becomes much less significant when
the two filters are both wide (filter set $b$) or both narrow (filter set
$c$).  Filter sets $d$ and $e$ show that the nonlinearity becomes more
prominent when the two filters with significantly different widths
are closer to each other.  This is because the dependence of $\lambda_A$
ratio on $A_0$ is larger when the two filters are closer to each other,
and this supports our interpretation of the nonlinear reddening vector of
the filter set F160W \& F222M.

\begin{figure}
%Fig 23
\epsscale{1.0}
\plotone{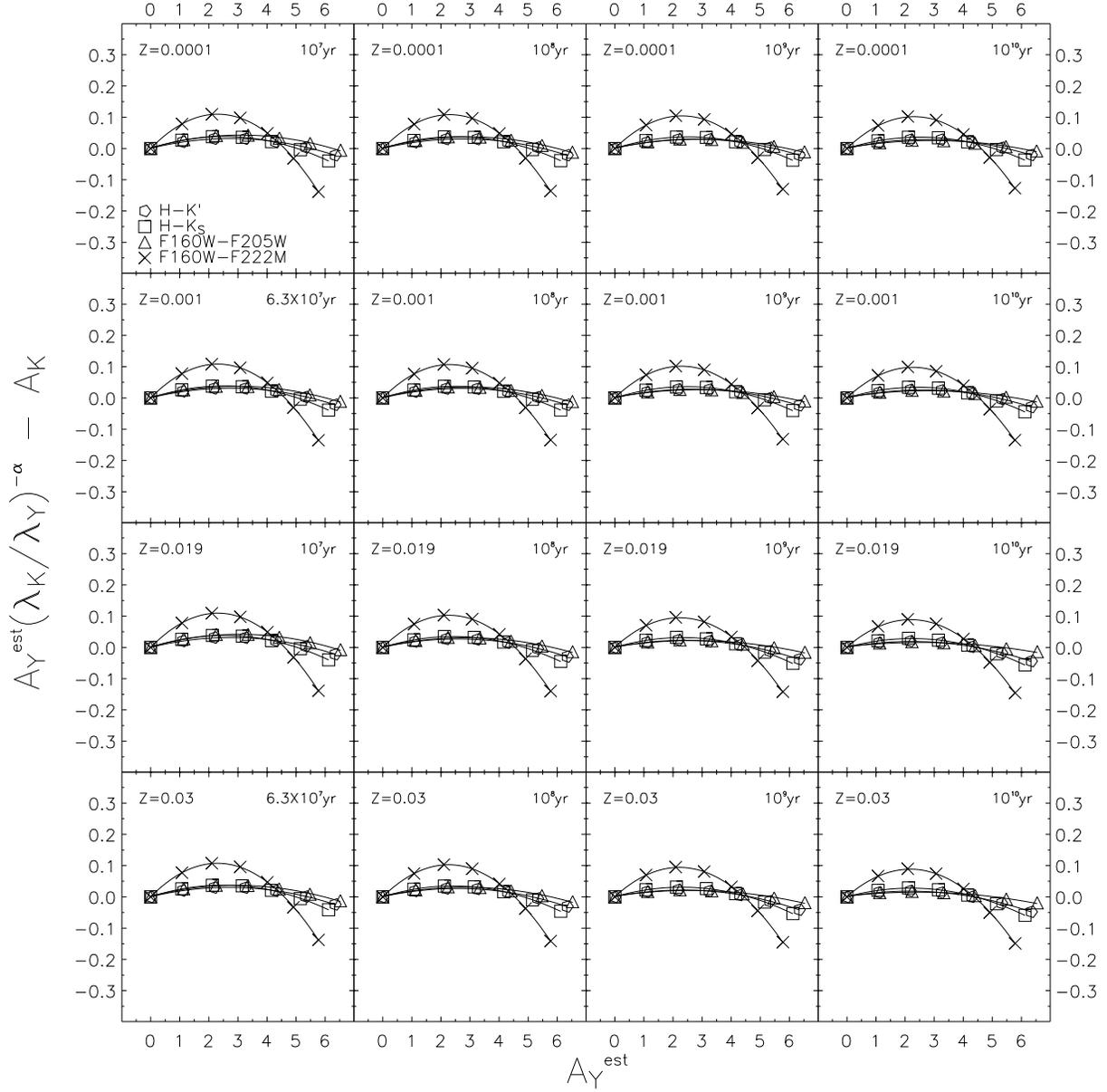}
\caption
{\label{fig:adiff4}The difference between the extinction values that are
estimated by equation~(\ref{A_est}) with our $H-K'$, $H-K_s$, F160W$-$F205W,
\& F160W$-$F222M colors and then converted to an extinction value at $K$
by equation~(\ref{A_trans}) with $\alpha=1.61$, and the extinction values
for $K$ that are estimated by equation~(\ref{A_est}) with our $H-K$ colors.
}
\end{figure}

Now we discuss how one may transform an extinction value
estimated for one filter in the $K$-band to the extinction value for
another filter.  When the empirical extinction law from observations is
not avaialable for the two filters, transformation of extinction values
between filters $Y$ and $Y'$ is normally obtained by
\begin{equation}
\label{A_trans}
	A_{Y'} = A_Y \left ( \frac{\lambda_{Y'}}{\lambda_Y} \right )^{-\alpha},
\end{equation}
assuming that the extinction has a functional form of a power-law, and
the filter transmission functions are Dirac delta functions
centered at their central wavelengths.  We find that once $A_Y$ is estimated
with $\alpha_{eff}$, equation~(\ref{A_trans}) with the original $\alpha$
value, 1.61, gives good estimates for $A_{Y'}$.  Figure~\ref{fig:adiff4}
shows the difference between the extinction values that are estimated by
equation~(\ref{A_est}) with our $H-K'$, $H-K_s$, F160W$-$F205W, \&
F160W$-$F222M colors and then converted to an extinction value at $K$
by equation~(\ref{A_trans}) with $\alpha=1.61$, and the actual extinction
values for $K$.  Except for filter set F160W$-$F222M again, extinction
values can be reliably transformed between different filter sets following
the above procedure.  It is interesting to note that the original $\alpha$
value of 1.61 is not suitable for estimating the extinction, but can be
used for transformation between appropriately estimated extinction values.

\section{SUMMARY}
\label{sec:summary}

We have calculated in a consistent way five near-infrared theoretical
isochrones for filter sets $H-K$, $H-K'$, $H-K_s$, F160W$-$F205W, and
F160W$-$F222M.  We presented isochrones for $Z$ of 0.0001 to 0.03, and
age of $10^7$ to $10^{10}$~yr.  Even in the same Vega magnitude system,
near-infrared colors of the same isochrone can be different by up to 0.18 mag
at the bright end of the isochrone for different filter sets.
The difference of intrinsic colors for a red giant between atmospheric
filters and the space filters is generally $\sim$0.1-0.15 mag.
We provided magnitude transformations between $K$-band filters
as a function of color from $H$ \& $K$ band filters.
We also presented isochrones with $A_K$ of up to 6 mag.
Isochrones for larger extinction values are found to be more vertically
straight.  We found that care is needed when comparing extinction
values that are estimated using different filter sets, in particular
when comparing those between atmospheric and space filter sets:
extinction values inferred using space filters can be in error 
by up to 0.3 mag.  To alleviate this problem, we introduced an
``effective extinction slope'' for each filter set and isochrone model,
which describes the extinction-dependent behaviour of isochrones in the
observed CMD.

\acknowledgements
S.S.K. thanks Hwankyung Sung for helpful discussion. We thank Jeff Valenti for
identifying H$^-$ as a source of opacity in cool stars.  We thank the anonymous
referee whose suggestions and comments significantly improved the content of
the present paper.  This work was supported by the research fund from Kyung
Hee University.  S.S.K. was supported by the Astrophysical Research Center for
the Structure and Evolution of the Cosmos (ARCSEC) of Korea Science
and Engineering Foundation through the Science Research Center (SRC) program.
M.G.L. was in part supported by the ABRL (R14-2002-058-01000-0) and the BK21
program.

%%%%%%%%%%%%%%%%%%%%%%%%%%%%%%%%%%%%%%%%%%%%%%%%%%%%%%%%%%%%%%%%%%%%%%%%%%%%%%%%

%%%%%%%%%%%%%%%%%%%%%%%%%%%%%%%%%%%%%%%%%%%%%%%%%%%%%%%%%%%%%%%%%%%%%%%%%%%%%%%%
\clearpage
%Table 1
\begin{deluxetable}{cclrrrcc}
\tabletypesize{\scriptsize}
\tablecolumns{8}
\tablewidth{0pt}
\tablecaption{
\label{table:trans1}Best-Fit Coefficients for Magnitude
Differences\tablenotemark{a} ($K < 4$~mag)}
\tablehead{
\colhead{} &
\colhead{Magnitude} &
\colhead{} &
\colhead{} &
\colhead{} &
\colhead{} &
\colhead{Residual\tablenotemark{b}} &
\colhead{Fitting Range\tablenotemark{c}} \\
\colhead{Color} &
\colhead{Difference} &
\colhead{$Z$} &
\colhead{$c_0$} &
\colhead{$c_1$} &
\colhead{$c_2$} &
\colhead{(mag)} &
\colhead{(mag $\sim$ mag)}
}
\startdata
$H - K$  & $K'$ $-$$K$  & 0.0001 & $-0.000$ & $ 0.092$ & $-0.166$ & $0.003$ & $-0.108 \sim 0.118$ \\
$H - K$  & $K'$ $-$$K$  & 0.001  & $ 0.004$ & $ 0.018$ & $-3.318$ & $0.009$ & $-0.072 \sim 0.087$ \\
$H - K$  & $K'$ $-$$K$  & 0.019  & $-0.001$ & $-0.100$ & $-2.285$ & $0.019$ & $-0.096 \sim 0.135$ \\
$H - K$  & $K'$ $-$$K$  & 0.03   & $-0.003$ & $-0.108$ & $-1.213$ & $0.014$ & $-0.057 \sim 0.168$ \\
$H - K$  & $K_s$$-$$K$  & 0.0001 & $ 0.000$ & $ 0.039$ & $-0.074$ & $0.003$ & $-0.108 \sim 0.118$ \\
$H - K$  & $K_s$$-$$K$  & 0.001  & $ 0.004$ & $-0.021$ & $-2.580$ & $0.008$ & $-0.072 \sim 0.087$ \\
$H - K$  & $K_s$$-$$K$  & 0.019  & $-0.001$ & $-0.130$ & $-1.851$ & $0.016$ & $-0.096 \sim 0.135$ \\
$H - K$  & $K_s$$-$$K$  & 0.03   & $-0.002$ & $-0.141$ & $-0.942$ & $0.012$ & $-0.057 \sim 0.168$ \\
$H - K$  & F205W$-$$K$  & 0.0001 & $-0.029$ & $ 0.183$ & $-0.145$ & $0.004$ & $-0.108 \sim 0.118$ \\
$H - K$  & F205W$-$$K$  & 0.001  & $-0.026$ & $ 0.134$ & $-2.389$ & $0.006$ & $-0.072 \sim 0.087$ \\
$H - K$  & F205W$-$$K$  & 0.019  & $-0.028$ & $ 0.060$ & $-1.843$ & $0.013$ & $-0.096 \sim 0.135$ \\
$H - K$  & F205W$-$$K$  & 0.03   & $-0.029$ & $ 0.070$ & $-1.035$ & $0.010$ & $-0.057 \sim 0.168$ \\
$H - K$  & F222M$-$$K$  & 0.0001 & $-0.031$ & $-0.042$ & $-0.394$ & $0.005$ & $-0.108 \sim 0.118$ \\
$H - K$  & F222M$-$$K$  & 0.001  & $-0.027$ & $-0.113$ & $-3.384$ & $0.008$ & $-0.072 \sim 0.087$ \\
$H - K$  & F222M$-$$K$  & 0.019  & $-0.034$ & $-0.240$ & $-2.119$ & $0.018$ & $-0.096 \sim 0.135$ \\
$H - K$  & F222M$-$$K$  & 0.03   & $-0.035$ & $-0.271$ & $-0.852$ & $0.014$ & $-0.057 \sim 0.168$ \\
$H - K'$  & $K$ $-$$K'$  & 0.0001 & $ 0.000$ & $-0.102$ & $ 0.228$ & $0.003$ & $-0.096 \sim 0.108$ \\
$H - K'$  & $K$ $-$$K'$  & 0.001  & $-0.003$ & $-0.057$ & $ 2.711$ & $0.006$ & $-0.065 \sim 0.110$ \\
$H - K'$  & $K$ $-$$K'$  & 0.019  & $ 0.002$ & $ 0.061$ & $ 1.356$ & $0.015$ & $-0.086 \sim 0.175$ \\
$H - K'$  & $K$ $-$$K'$  & 0.03   & $ 0.002$ & $ 0.105$ & $ 0.639$ & $0.011$ & $-0.050 \sim 0.209$ \\
$H - K'$  & $K_s$$-$$K'$  & 0.0001 & $ 0.001$ & $-0.059$ & $ 0.124$ & $0.003$ & $-0.096 \sim 0.108$ \\
$H - K'$  & $K_s$$-$$K'$  & 0.001  & $-0.000$ & $-0.048$ & $ 0.661$ & $0.002$ & $-0.065 \sim 0.110$ \\
$H - K'$  & $K_s$$-$$K'$  & 0.019  & $ 0.000$ & $-0.034$ & $ 0.321$ & $0.003$ & $-0.086 \sim 0.175$ \\
$H - K'$  & $K_s$$-$$K'$  & 0.03   & $ 0.000$ & $-0.029$ & $ 0.184$ & $0.003$ & $-0.050 \sim 0.209$ \\
$H - K'$  & F205W$-$$K'$  & 0.0001 & $-0.029$ & $ 0.100$ & $ 0.008$ & $0.003$ & $-0.096 \sim 0.108$ \\
$H - K'$  & F205W$-$$K'$  & 0.001  & $-0.029$ & $ 0.111$ & $ 0.467$ & $0.003$ & $-0.065 \sim 0.110$ \\
$H - K'$  & F205W$-$$K'$  & 0.019  & $-0.027$ & $ 0.147$ & $ 0.053$ & $0.004$ & $-0.086 \sim 0.175$ \\
$H - K'$  & F205W$-$$K'$  & 0.03   & $-0.027$ & $ 0.165$ & $-0.031$ & $0.004$ & $-0.050 \sim 0.209$ \\
$H - K'$  & F222M$-$$K'$  & 0.0001 & $-0.031$ & $-0.147$ & $-0.240$ & $0.005$ & $-0.096 \sim 0.108$ \\
$H - K'$  & F222M$-$$K'$  & 0.001  & $-0.032$ & $-0.142$ & $ 0.409$ & $0.005$ & $-0.065 \sim 0.110$ \\
$H - K'$  & F222M$-$$K'$  & 0.019  & $-0.033$ & $-0.140$ & $ 0.345$ & $0.005$ & $-0.086 \sim 0.175$ \\
$H - K'$  & F222M$-$$K'$  & 0.03   & $-0.032$ & $-0.156$ & $ 0.381$ & $0.005$ & $-0.050 \sim 0.209$ \\
$H - K_s$ & $K$ $-$$K_s$ & 0.0001 & $-0.000$ & $-0.041$ & $ 0.086$ & $0.003$ & $-0.104 \sim 0.114$ \\
$H - K_s$ & $K$ $-$$K_s$ & 0.001  & $-0.003$ & $-0.000$ & $ 2.027$ & $0.005$ & $-0.070 \sim 0.108$ \\
$H - K_s$ & $K$ $-$$K_s$ & 0.019  & $ 0.001$ & $ 0.095$ & $ 1.075$ & $0.013$ & $-0.092 \sim 0.174$ \\
$H - K_s$ & $K$ $-$$K_s$ & 0.03   & $ 0.002$ & $ 0.127$ & $ 0.500$ & $0.009$ & $-0.055 \sim 0.209$ \\
$H - K_s$ & $K'$ $-$$K_s$ & 0.0001 & $-0.001$ & $ 0.055$ & $-0.103$ & $0.002$ & $-0.104 \sim 0.114$ \\
$H - K_s$ & $K'$ $-$$K_s$ & 0.001  & $ 0.000$ & $ 0.044$ & $-0.633$ & $0.002$ & $-0.070 \sim 0.108$ \\
$H - K_s$ & $K'$ $-$$K_s$ & 0.019  & $-0.000$ & $ 0.032$ & $-0.314$ & $0.003$ & $-0.092 \sim 0.174$ \\
$H - K_s$ & $K'$ $-$$K_s$ & 0.03   & $-0.000$ & $ 0.029$ & $-0.184$ & $0.003$ & $-0.055 \sim 0.209$ \\
$H - K_s$ & F205W$-$$K_s$ & 0.0001 & $-0.029$ & $ 0.150$ & $-0.087$ & $0.005$ & $-0.104 \sim 0.114$ \\
$H - K_s$ & F205W$-$$K_s$ & 0.001  & $-0.029$ & $ 0.151$ & $-0.106$ & $0.003$ & $-0.070 \sim 0.108$ \\
$H - K_s$ & F205W$-$$K_s$ & 0.019  & $-0.028$ & $ 0.174$ & $-0.209$ & $0.003$ & $-0.092 \sim 0.174$ \\
$H - K_s$ & F205W$-$$K_s$ & 0.03   & $-0.028$ & $ 0.188$ & $-0.179$ & $0.003$ & $-0.055 \sim 0.209$ \\
$H - K_s$ & F222M$-$$K_s$ & 0.0001 & $-0.032$ & $-0.084$ & $-0.335$ & $0.006$ & $-0.104 \sim 0.114$ \\
$H - K_s$ & F222M$-$$K_s$ & 0.001  & $-0.032$ & $-0.089$ & $-0.336$ & $0.006$ & $-0.070 \sim 0.108$ \\
$H - K_s$ & F222M$-$$K_s$ & 0.019  & $-0.033$ & $-0.100$ & $-0.021$ & $0.007$ & $-0.092 \sim 0.174$ \\
$H - K_s$ & F222M$-$$K_s$ & 0.03   & $-0.033$ & $-0.121$ & $ 0.163$ & $0.007$ & $-0.055 \sim 0.209$ \\
F160W$-$F205W & $K$ $-$F205W & 0.0001 & $ 0.030$ & $-0.164$ & $ 0.284$ & $0.007$ & $-0.098 \sim 0.154$ \\
F160W$-$F205W & $K$ $-$F205W & 0.001  & $ 0.028$ & $-0.168$ & $ 1.267$ & $0.004$ & $-0.067 \sim 0.169$ \\
F160W$-$F205W & $K$ $-$F205W & 0.019  & $ 0.030$ & $-0.084$ & $ 0.681$ & $0.012$ & $-0.087 \sim 0.261$ \\
F160W$-$F205W & $K$ $-$F205W & 0.03   & $ 0.030$ & $-0.055$ & $ 0.410$ & $0.009$ & $-0.052 \sim 0.279$ \\
F160W$-$F205W & $K'$ $-$F205W & 0.0001 & $ 0.029$ & $-0.077$ & $ 0.064$ & $0.004$ & $-0.098 \sim 0.154$ \\
F160W$-$F205W & $K'$ $-$F205W & 0.001  & $ 0.029$ & $-0.069$ & $-0.211$ & $0.004$ & $-0.067 \sim 0.169$ \\
F160W$-$F205W & $K'$ $-$F205W & 0.019  & $ 0.029$ & $-0.111$ & $ 0.018$ & $0.005$ & $-0.087 \sim 0.261$ \\
F160W$-$F205W & $K'$ $-$F205W & 0.03   & $ 0.028$ & $-0.117$ & $-0.019$ & $0.004$ & $-0.052 \sim 0.279$ \\
F160W$-$F205W & $K_s$$-$F205W & 0.0001 & $ 0.030$ & $-0.126$ & $ 0.186$ & $0.007$ & $-0.098 \sim 0.154$ \\
F160W$-$F205W & $K_s$$-$F205W & 0.001  & $ 0.030$ & $-0.117$ & $ 0.158$ & $0.005$ & $-0.067 \sim 0.169$ \\
F160W$-$F205W & $K_s$$-$F205W & 0.019  & $ 0.029$ & $-0.141$ & $ 0.185$ & $0.004$ & $-0.087 \sim 0.261$ \\
F160W$-$F205W & $K_s$$-$F205W & 0.03   & $ 0.029$ & $-0.141$ & $ 0.089$ & $0.005$ & $-0.052 \sim 0.279$ \\
F160W$-$F205W & F222M$-$F205W & 0.0001 & $-0.001$ & $-0.187$ & $ 0.021$ & $0.005$ & $-0.098 \sim 0.154$ \\
F160W$-$F205W & F222M$-$F205W & 0.001  & $-0.001$ & $-0.185$ & $ 0.100$ & $0.002$ & $-0.067 \sim 0.169$ \\
F160W$-$F205W & F222M$-$F205W & 0.019  & $-0.003$ & $-0.225$ & $ 0.248$ & $0.005$ & $-0.087 \sim 0.261$ \\
F160W$-$F205W & F222M$-$F205W & 0.03   & $-0.002$ & $-0.238$ & $ 0.198$ & $0.003$ & $-0.052 \sim 0.279$ \\
F160W$-$F222M & $K$ $-$F222M & 0.0001 & $ 0.031$ & $ 0.020$ & $ 0.180$ & $0.005$ & $-0.118 \sim 0.188$ \\
F160W$-$F222M & $K$ $-$F222M & 0.001  & $ 0.029$ & $ 0.013$ & $ 0.847$ & $0.005$ & $-0.079 \sim 0.199$ \\
F160W$-$F222M & $K$ $-$F222M & 0.019  & $ 0.032$ & $ 0.110$ & $ 0.358$ & $0.011$ & $-0.105 \sim 0.305$ \\
F160W$-$F222M & $K$ $-$F222M & 0.03   & $ 0.032$ & $ 0.143$ & $ 0.186$ & $0.010$ & $-0.061 \sim 0.333$ \\
F160W$-$F222M & $K'$ $-$F222M & 0.0001 & $ 0.030$ & $ 0.092$ & $ 0.035$ & $0.003$ & $-0.118 \sim 0.188$ \\
F160W$-$F222M & $K'$ $-$F222M & 0.001  & $ 0.031$ & $ 0.099$ & $-0.218$ & $0.003$ & $-0.079 \sim 0.199$ \\
F160W$-$F222M & $K'$ $-$F222M & 0.019  & $ 0.031$ & $ 0.095$ & $-0.155$ & $0.003$ & $-0.105 \sim 0.305$ \\
F160W$-$F222M & $K'$ $-$F222M & 0.03   & $ 0.031$ & $ 0.100$ & $-0.145$ & $0.003$ & $-0.061 \sim 0.333$ \\
F160W$-$F222M & $K_s$$-$F222M & 0.0001 & $ 0.031$ & $ 0.051$ & $ 0.114$ & $0.005$ & $-0.118 \sim 0.188$ \\
F160W$-$F222M & $K_s$$-$F222M & 0.001  & $ 0.031$ & $ 0.058$ & $ 0.047$ & $0.005$ & $-0.079 \sim 0.199$ \\
F160W$-$F222M & $K_s$$-$F222M & 0.019  & $ 0.032$ & $ 0.069$ & $-0.032$ & $0.005$ & $-0.105 \sim 0.305$ \\
F160W$-$F222M & $K_s$$-$F222M & 0.03   & $ 0.031$ & $ 0.079$ & $-0.067$ & $0.005$ & $-0.061 \sim 0.333$ \\
F160W$-$F222M & F205W$-$F222M & 0.0001 & $ 0.000$ & $ 0.157$ & $-0.003$ & $0.004$ & $-0.118 \sim 0.188$ \\
F160W$-$F222M & F205W$-$F222M & 0.001  & $ 0.001$ & $ 0.157$ & $-0.061$ & $0.002$ & $-0.079 \sim 0.199$ \\
F160W$-$F222M & F205W$-$F222M & 0.019  & $ 0.002$ & $ 0.185$ & $-0.150$ & $0.005$ & $-0.105 \sim 0.305$ \\
F160W$-$F222M & F205W$-$F222M & 0.03   & $ 0.002$ & $ 0.194$ & $-0.115$ & $0.003$ & $-0.061 \sim 0.333$ \\
\enddata
\tablecomments{Only the data points that have $\log T_{eff} \ge
3500$~K and $\log g \ge 0$ were considered for the fitting.}
\tablenotetext{a}{Magnitude differences are fitted to a function
$[{\rm Mag\, Diff}] = c_0 + c_1[{\rm Color}] + c_2[{\rm Color}]^2$.}
\tablenotetext{b}{The largest absolute residual.}
\tablenotetext{c}{Color range where the fit is valid.}
\end{deluxetable}

\clearpage
%Table 2
\begin{deluxetable}{cclrrcc}
\tabletypesize{\scriptsize}
\tablecolumns{7}
\tablewidth{0pt}
\tablecaption{
\label{table:trans2}Best-Fit Coefficients for Magnitude
Differences\tablenotemark{a} ($K > 4$~mag)}
\tablehead{
\colhead{} &
\colhead{Magnitude} &
\colhead{} &
\colhead{} &
\colhead{} &
\colhead{Residual\tablenotemark{b}} &
\colhead{Fitting Range\tablenotemark{c}} \\
\colhead{Color} &
\colhead{Difference} &
\colhead{$Z$} &
\colhead{$c_0$} &
\colhead{$c_1$} &
\colhead{(mag)} &
\colhead{(mag $\sim$ mag)}
}
\startdata
$H - K$  & $K'$ $-$$K$  & 0.0001 & $ 0.001$ & $ 0.112$ & $0.001$ & $ 0.026 \sim 0.195$ \\
$H - K$  & $K'$ $-$$K$  & 0.001  & $-0.001$ & $ 0.097$ & $0.003$ & $ 0.028 \sim 0.223$ \\
$H - K$  & $K'$ $-$$K$  & 0.019  & $-0.014$ & $ 0.131$ & $0.004$ & $ 0.046 \sim 0.242$ \\
$H - K$  & $K'$ $-$$K$  & 0.03   & $-0.017$ & $ 0.136$ & $0.005$ & $ 0.050 \sim 0.235$ \\
$H - K$  & $K_s$$-$$K$  & 0.0001 & $ 0.002$ & $ 0.053$ & $0.001$ & $ 0.026 \sim 0.195$ \\
$H - K$  & $K_s$$-$$K$  & 0.001  & $ 0.001$ & $ 0.029$ & $0.002$ & $ 0.028 \sim 0.223$ \\
$H - K$  & $K_s$$-$$K$  & 0.019  & $-0.011$ & $ 0.060$ & $0.003$ & $ 0.046 \sim 0.242$ \\
$H - K$  & $K_s$$-$$K$  & 0.03   & $-0.013$ & $ 0.060$ & $0.004$ & $ 0.050 \sim 0.235$ \\
$H - K$  & F205W$-$$K$  & 0.0001 & $-0.027$ & $ 0.210$ & $0.001$ & $ 0.026 \sim 0.195$ \\
$H - K$  & F205W$-$$K$  & 0.001  & $-0.030$ & $ 0.198$ & $0.003$ & $ 0.028 \sim 0.223$ \\
$H - K$  & F205W$-$$K$  & 0.019  & $-0.041$ & $ 0.232$ & $0.004$ & $ 0.046 \sim 0.242$ \\
$H - K$  & F205W$-$$K$  & 0.03   & $-0.042$ & $ 0.233$ & $0.004$ & $ 0.050 \sim 0.235$ \\
$H - K$  & F222M$-$$K$  & 0.0001 & $-0.032$ & $-0.050$ & $0.001$ & $ 0.026 \sim 0.195$ \\
$H - K$  & F222M$-$$K$  & 0.001  & $-0.033$ & $-0.066$ & $0.003$ & $ 0.028 \sim 0.223$ \\
$H - K$  & F222M$-$$K$  & 0.019  & $-0.046$ & $-0.037$ & $0.003$ & $ 0.046 \sim 0.242$ \\
$H - K$  & F222M$-$$K$  & 0.03   & $-0.048$ & $-0.038$ & $0.003$ & $ 0.050 \sim 0.235$ \\
$H - K'$  & $K$ $-$$K'$  & 0.0001 & $-0.001$ & $-0.126$ & $0.001$ & $ 0.022 \sim 0.172$ \\
$H - K'$  & $K$ $-$$K'$  & 0.001  & $ 0.001$ & $-0.106$ & $0.003$ & $ 0.024 \sim 0.201$ \\
$H - K'$  & $K$ $-$$K'$  & 0.019  & $ 0.017$ & $-0.149$ & $0.005$ & $ 0.050 \sim 0.225$ \\
$H - K'$  & $K$ $-$$K'$  & 0.03   & $ 0.020$ & $-0.160$ & $0.004$ & $ 0.058 \sim 0.219$ \\
$H - K'$  & $K_s$$-$$K'$  & 0.0001 & $ 0.000$ & $-0.066$ & $0.001$ & $ 0.022 \sim 0.172$ \\
$H - K'$  & $K_s$$-$$K'$  & 0.001  & $ 0.002$ & $-0.074$ & $0.002$ & $ 0.024 \sim 0.201$ \\
$H - K'$  & $K_s$$-$$K'$  & 0.019  & $ 0.004$ & $-0.081$ & $0.001$ & $ 0.050 \sim 0.225$ \\
$H - K'$  & $K_s$$-$$K'$  & 0.03   & $ 0.005$ & $-0.089$ & $0.001$ & $ 0.058 \sim 0.219$ \\
$H - K'$  & F205W$-$$K'$  & 0.0001 & $-0.028$ & $ 0.110$ & $0.001$ & $ 0.022 \sim 0.172$ \\
$H - K'$  & F205W$-$$K'$  & 0.001  & $-0.029$ & $ 0.111$ & $0.001$ & $ 0.024 \sim 0.201$ \\
$H - K'$  & F205W$-$$K'$  & 0.019  & $-0.028$ & $ 0.116$ & $0.001$ & $ 0.050 \sim 0.225$ \\
$H - K'$  & F205W$-$$K'$  & 0.03   & $-0.027$ & $ 0.112$ & $0.001$ & $ 0.058 \sim 0.219$ \\
$H - K'$  & F222M$-$$K'$  & 0.0001 & $-0.033$ & $-0.182$ & $0.001$ & $ 0.022 \sim 0.172$ \\
$H - K'$  & F222M$-$$K'$  & 0.001  & $-0.032$ & $-0.180$ & $0.002$ & $ 0.024 \sim 0.201$ \\
$H - K'$  & F222M$-$$K'$  & 0.019  & $-0.028$ & $-0.193$ & $0.002$ & $ 0.050 \sim 0.225$ \\
$H - K'$  & F222M$-$$K'$  & 0.03   & $-0.028$ & $-0.203$ & $0.002$ & $ 0.058 \sim 0.219$ \\
$H - K_s$ & $K$ $-$$K_s$ & 0.0001 & $-0.002$ & $-0.056$ & $0.001$ & $ 0.024 \sim 0.184$ \\
$H - K_s$ & $K$ $-$$K_s$ & 0.001  & $-0.001$ & $-0.030$ & $0.002$ & $ 0.025 \sim 0.215$ \\
$H - K_s$ & $K$ $-$$K_s$ & 0.019  & $ 0.012$ & $-0.063$ & $0.004$ & $ 0.051 \sim 0.239$ \\
$H - K_s$ & $K$ $-$$K_s$ & 0.03   & $ 0.014$ & $-0.065$ & $0.004$ & $ 0.059 \sim 0.234$ \\
$H - K_s$ & $K'$ $-$$K_s$ & 0.0001 & $-0.000$ & $ 0.062$ & $0.001$ & $ 0.024 \sim 0.184$ \\
$H - K_s$ & $K'$ $-$$K_s$ & 0.001  & $-0.002$ & $ 0.069$ & $0.002$ & $ 0.025 \sim 0.215$ \\
$H - K_s$ & $K'$ $-$$K_s$ & 0.019  & $-0.004$ & $ 0.075$ & $0.001$ & $ 0.051 \sim 0.239$ \\
$H - K_s$ & $K'$ $-$$K_s$ & 0.03   & $-0.005$ & $ 0.082$ & $0.001$ & $ 0.059 \sim 0.234$ \\
$H - K_s$ & F205W$-$$K_s$ & 0.0001 & $-0.028$ & $ 0.166$ & $0.001$ & $ 0.024 \sim 0.184$ \\
$H - K_s$ & F205W$-$$K_s$ & 0.001  & $-0.030$ & $ 0.173$ & $0.002$ & $ 0.025 \sim 0.215$ \\
$H - K_s$ & F205W$-$$K_s$ & 0.019  & $-0.032$ & $ 0.182$ & $0.001$ & $ 0.051 \sim 0.239$ \\
$H - K_s$ & F205W$-$$K_s$ & 0.03   & $-0.031$ & $ 0.184$ & $0.001$ & $ 0.059 \sim 0.234$ \\
$H - K_s$ & F222M$-$$K_s$ & 0.0001 & $-0.034$ & $-0.109$ & $0.001$ & $ 0.024 \sim 0.184$ \\
$H - K_s$ & F222M$-$$K_s$ & 0.001  & $-0.034$ & $-0.098$ & $0.001$ & $ 0.025 \sim 0.215$ \\
$H - K_s$ & F222M$-$$K_s$ & 0.019  & $-0.033$ & $-0.103$ & $0.001$ & $ 0.051 \sim 0.239$ \\
$H - K_s$ & F222M$-$$K_s$ & 0.03   & $-0.034$ & $-0.105$ & $0.001$ & $ 0.059 \sim 0.234$ \\
F160W$-$F205W & $K$ $-$F205W & 0.0001 & $ 0.032$ & $-0.217$ & $0.001$ & $ 0.047 \sim 0.211$ \\
F160W$-$F205W & $K$ $-$F205W & 0.001  & $ 0.036$ & $-0.194$ & $0.005$ & $ 0.054 \sim 0.246$ \\
F160W$-$F205W & $K$ $-$F205W & 0.019  & $ 0.059$ & $-0.269$ & $0.009$ & $ 0.092 \sim 0.274$ \\
F160W$-$F205W & $K$ $-$F205W & 0.03   & $ 0.063$ & $-0.277$ & $0.009$ & $ 0.102 \sim 0.269$ \\
F160W$-$F205W & $K'$ $-$F205W & 0.0001 & $ 0.030$ & $-0.101$ & $0.001$ & $ 0.047 \sim 0.211$ \\
F160W$-$F205W & $K'$ $-$F205W & 0.001  & $ 0.032$ & $-0.100$ & $0.003$ & $ 0.054 \sim 0.246$ \\
F160W$-$F205W & $K'$ $-$F205W & 0.019  & $ 0.035$ & $-0.118$ & $0.002$ & $ 0.092 \sim 0.274$ \\
F160W$-$F205W & $K'$ $-$F205W & 0.03   & $ 0.034$ & $-0.114$ & $0.002$ & $ 0.102 \sim 0.269$ \\
F160W$-$F205W & $K_s$$-$F205W & 0.0001 & $ 0.032$ & $-0.163$ & $0.001$ & $ 0.047 \sim 0.211$ \\
F160W$-$F205W & $K_s$$-$F205W & 0.001  & $ 0.036$ & $-0.166$ & $0.004$ & $ 0.054 \sim 0.246$ \\
F160W$-$F205W & $K_s$$-$F205W & 0.019  & $ 0.044$ & $-0.200$ & $0.004$ & $ 0.092 \sim 0.274$ \\
F160W$-$F205W & $K_s$$-$F205W & 0.03   & $ 0.044$ & $-0.206$ & $0.004$ & $ 0.102 \sim 0.269$ \\
F160W$-$F205W & F222M$-$F205W & 0.0001 & $ 0.001$ & $-0.269$ & $0.001$ & $ 0.047 \sim 0.211$ \\
F160W$-$F205W & F222M$-$F205W & 0.001  & $ 0.005$ & $-0.260$ & $0.005$ & $ 0.054 \sim 0.246$ \\
F160W$-$F205W & F222M$-$F205W & 0.019  & $ 0.017$ & $-0.314$ & $0.006$ & $ 0.092 \sim 0.274$ \\
F160W$-$F205W & F222M$-$F205W & 0.03   & $ 0.018$ & $-0.322$ & $0.007$ & $ 0.102 \sim 0.269$ \\
F160W$-$F222M & $K$ $-$F222M & 0.0001 & $ 0.031$ & $ 0.041$ & $0.001$ & $ 0.058 \sim 0.267$ \\
F160W$-$F222M & $K$ $-$F222M & 0.001  & $ 0.031$ & $ 0.052$ & $0.002$ & $ 0.066 \sim 0.309$ \\
F160W$-$F222M & $K$ $-$F222M & 0.019  & $ 0.043$ & $ 0.034$ & $0.003$ & $ 0.110 \sim 0.344$ \\
F160W$-$F222M & $K$ $-$F222M & 0.03   & $ 0.046$ & $ 0.034$ & $0.002$ & $ 0.124 \sim 0.339$ \\
F160W$-$F222M & $K'$ $-$F222M & 0.0001 & $ 0.029$ & $ 0.132$ & $0.001$ & $ 0.058 \sim 0.267$ \\
F160W$-$F222M & $K'$ $-$F222M & 0.001  & $ 0.028$ & $ 0.128$ & $0.003$ & $ 0.066 \sim 0.309$ \\
F160W$-$F222M & $K'$ $-$F222M & 0.019  & $ 0.020$ & $ 0.150$ & $0.004$ & $ 0.110 \sim 0.344$ \\
F160W$-$F222M & $K'$ $-$F222M & 0.03   & $ 0.019$ & $ 0.158$ & $0.004$ & $ 0.124 \sim 0.339$ \\
F160W$-$F222M & $K_s$$-$F222M & 0.0001 & $ 0.031$ & $ 0.084$ & $0.001$ & $ 0.058 \sim 0.267$ \\
F160W$-$F222M & $K_s$$-$F222M & 0.001  & $ 0.031$ & $ 0.075$ & $0.001$ & $ 0.066 \sim 0.309$ \\
F160W$-$F222M & $K_s$$-$F222M & 0.019  & $ 0.028$ & $ 0.087$ & $0.002$ & $ 0.110 \sim 0.344$ \\
F160W$-$F222M & $K_s$$-$F222M & 0.03   & $ 0.028$ & $ 0.089$ & $0.002$ & $ 0.124 \sim 0.339$ \\
F160W$-$F222M & F205W$-$F222M & 0.0001 & $-0.001$ & $ 0.212$ & $0.001$ & $ 0.058 \sim 0.267$ \\
F160W$-$F222M & F205W$-$F222M & 0.001  & $-0.004$ & $ 0.208$ & $0.004$ & $ 0.066 \sim 0.309$ \\
F160W$-$F222M & F205W$-$F222M & 0.019  & $-0.013$ & $ 0.240$ & $0.005$ & $ 0.110 \sim 0.344$ \\
F160W$-$F222M & F205W$-$F222M & 0.03   & $-0.014$ & $ 0.245$ & $0.005$ & $ 0.124 \sim 0.339$ \\
\enddata
\tablecomments{Only the data points that have $\log T_{eff} \ge
3500$~K and $\log g \ge 0$ were considered for the fitting.}
\tablenotetext{a}{Magnitude differences are fitted to a function
${\rm [Mag\, Diff]} = c_0 + c_1[{\rm Color}]$.}
\tablenotetext{b}{The largest absolute residual.}
\tablenotetext{c}{Color range where the fit is valid.}
\end{deluxetable}

\clearpage
%Table 3
\begin{deluxetable}{ccccccc}
\tablecolumns{7}
\tablewidth{0pt}
\tablecaption{
\label{table:lambdac}Central Wavelength $\lambda_c$ ($\mu$m)}
\tablehead{
\colhead{$H$} &
\colhead{$K$} &
\colhead{$K'$} &
\colhead{$K_s$} &
\colhead{F160W} &
\colhead{F205W} &
\colhead{F222M}
}
\startdata
1.646 &  2.212 &  2.114 &  2.160 &  1.610 &  2.079 &  2.219 \\
\enddata
\tablecomments{$\lambda_c$ is defined by equation (\ref{lambda_c}).}
\end{deluxetable}

\clearpage
%Table 4
\begin{deluxetable}{lcccccc}
\tabletypesize{\scriptsize}
\tablecolumns{7}
\tablewidth{0pt}
\tablecaption{
\label{table:alpha_eff}Averages \& Standard Deviations of $\alpha_{eff}$ Values}
\tablehead{
\multicolumn{2}{c}{Isochrone Model} &
\colhead{} &
\colhead{} &
\colhead{} &
\colhead{} &
\colhead{} \\ \cline{1-2}
\colhead{$Z$} &
\colhead{Age} &
\colhead{$H-K$} &
\colhead{$H-K'$} &
\colhead{$H-K_s$} &
\colhead{F160W$-$F205W} &
\colhead{F160W$-$F222M}
}
\startdata
  0.0001 &             $10^7$ &  1.574$\pm$0.001 &  1.572$\pm$0.001 &  1.574$\pm$0.001 &  1.544$\pm$0.001 &  1.566$\pm$0.001 \\
  0.0001 &             $10^8$ &  1.570$\pm$0.002 &  1.567$\pm$0.003 &  1.569$\pm$0.003 &  1.539$\pm$0.005 &  1.557$\pm$0.006 \\
  0.0001 &             $10^9$ &  1.566$\pm$0.002 &  1.563$\pm$0.002 &  1.565$\pm$0.002 &  1.528$\pm$0.008 &  1.544$\pm$0.008 \\
  0.0001 &          $10^{10}$ &  1.564$\pm$0.002 &  1.560$\pm$0.002 &  1.562$\pm$0.002 &  1.520$\pm$0.007 &  1.537$\pm$0.006 \\
  0.001  &  $6.3 \times 10^7$ &  1.570$\pm$0.003 &  1.567$\pm$0.003 &  1.570$\pm$0.003 &  1.539$\pm$0.005 &  1.557$\pm$0.007 \\
  0.001  &             $10^8$ &  1.568$\pm$0.004 &  1.565$\pm$0.003 &  1.568$\pm$0.004 &  1.536$\pm$0.009 &  1.553$\pm$0.009 \\
  0.001  &             $10^9$ &  1.562$\pm$0.005 &  1.560$\pm$0.003 &  1.563$\pm$0.003 &  1.520$\pm$0.009 &  1.538$\pm$0.006 \\
  0.001  &          $10^{10}$ &  1.556$\pm$0.009 &  1.557$\pm$0.005 &  1.559$\pm$0.005 &  1.511$\pm$0.013 &  1.532$\pm$0.009 \\
  0.019  &             $10^7$ &  1.574$\pm$0.001 &  1.571$\pm$0.001 &  1.574$\pm$0.001 &  1.545$\pm$0.001 &  1.566$\pm$0.001 \\
  0.019  &             $10^8$ &  1.561$\pm$0.013 &  1.562$\pm$0.008 &  1.564$\pm$0.009 &  1.526$\pm$0.021 &  1.547$\pm$0.017 \\
  0.019  &             $10^9$ &  1.549$\pm$0.009 &  1.555$\pm$0.006 &  1.555$\pm$0.006 &  1.505$\pm$0.015 &  1.529$\pm$0.010 \\
  0.019  &          $10^{10}$ &  1.539$\pm$0.010 &  1.548$\pm$0.007 &  1.548$\pm$0.007 &  1.489$\pm$0.016 &  1.517$\pm$0.012 \\
  0.03   &  $6.3 \times 10^7$ &  1.568$\pm$0.008 &  1.567$\pm$0.005 &  1.569$\pm$0.006 &  1.538$\pm$0.012 &  1.557$\pm$0.011 \\
  0.03   &             $10^8$ &  1.560$\pm$0.014 &  1.562$\pm$0.008 &  1.563$\pm$0.009 &  1.526$\pm$0.022 &  1.547$\pm$0.017 \\
  0.03   &             $10^9$ &  1.547$\pm$0.010 &  1.554$\pm$0.006 &  1.554$\pm$0.006 &  1.503$\pm$0.014 &  1.528$\pm$0.010 \\
  0.03   &          $10^{10}$ &  1.538$\pm$0.009 &  1.548$\pm$0.006 &  1.548$\pm$0.006 &  1.489$\pm$0.013 &  1.517$\pm$0.010 \\
\enddata
\tablecomments{Data are presented in the form of [Average $\pm$ Standard
Deviation].  The average and standard deviation values are calculated from
the data points of each isochrone whose intrinsic $K$ magnitudes are
between $-6$ and 0~mag.}
\end{deluxetable}

%%%%%%%%%%%%%%%%%%%%%%%%%%%%%%%%%%%%%%%%%%%%%%%%%%%%%%%%%%%%%%%%%%%%%%%%%%%%%%%%
\clearpage

\clearpage

\end{document}